\documentclass[twocolumn,aps,prb,superscriptaddress,groupedaddress]{revtex4}
\usepackage[latin9]{inputenc}
\usepackage{amsmath}
\usepackage{amssymb}
\usepackage{graphicx}
\usepackage{natbib}
\usepackage{color}
\usepackage{ulem}
\makeatletter

\providecommand{\tabularnewline}{\\}

\usepackage{dcolumn}
\usepackage{bm}

\newcommand{\add}[1]{#1}
\newcommand{\del}[1]{}

\def\em{\it} 
\renewcommand{\emph}[1]{\textit{#1}}

\newsavebox{\myfig}
\newsavebox{\mylbl}
\newlength{\myceil}
\newcommand{\lblgraphics}[3][white]
{%
	
	\savebox{\mylbl}{\large\color{#1}{\bf\,#2\vphantom{A}}}
	\savebox{\myfig}{#3}
	\setlength{\myceil}{\ht\myfig}
	\addtolength{\myceil}{-\ht\mylbl}
	{\hspace*{-\fontdimen2\font}{\ooalign{\usebox{\myfig}\cr \raisebox{\myceil}{\usebox{\mylbl}}}}}
}

\makeatother

\begin{document}

\title{Progress in structure recovery from low dose exposures: Mixed molecular
adsorption, exploitation of symmetry and reconstruction from the minimum
signal level}

\author{C.~Kramberger\textsuperscript{{*}}}

\author{J.C.~Meyer\textsuperscript{\dag{}}}

\affiliation{University of Vienna, Faculty of Physics, Strudlhofgass 4, 1090 Vienna}

\date{\today}
\begin{abstract}
We investigate the recovery of structures from large-area, low dose
exposures that distribute the dose over many identical copies of an
object. The reconstruction is done via a maximum likelihood approach
that does neither require to identify nor align the individual particles. 
We also simulate small molecular adsorbates on graphene and demonstrate the retrieval of images
with atomic resolution from \add{large area and} extremely low dose raw data. Doses as
low as 5 $e^{-}/$\AA$^{2}$ are sufficient if all symmetries (translations,
rotations and mirrors) of the supporting membrane are exploited to
retrieve the structure of individual adsorbed molecules. We compare
different optimization schemes, consider mixed molecules and adsorption
sites, and requirements on the amount of data. We further demonstrate
that the maximum likelihood approach is only count limited by requiring
at least three independent counts per entity. \add{Finally, we demonstrate that the approach works with real experiental data and in presence of aberrations.}

\textit{\vspace*{1cm}
}

\emph{keywords: Radiation damage, Electron microscopy, Maximum likelihood
reconstruction}

\emph{\textsuperscript{\textit{{*}}}email: Christian.Kramberger-Kaplan@univie.ac.at}

\emph{\textsuperscript{\textit{\dag{}}}email: Jannik.Meyer@univie.ac.at}
\end{abstract}
\maketitle

\section{Introduction}

With the advent of aberration corrected electron microscopy \cite{Haider1998,Krivanek1999},
individual atoms of light elements can be directly imaged in atomic
resolution \cite{Suenaga2007,Gass2008,Gomez-Navarro2010,Krivanek2010,Suenaga2012}
using electron energies below the knock-on damage threshold \cite{Banhart1999,Meyer2012}.
However, this only applies to certain materials, among organic materials
most notably graphene or carbon nanotubes are not affected by other
mechanisms of beam damage. The relative irradiation resilience of
low dimensional carbon allotropes is strongly contrasted by virtually
any other organic materials or molecules that are rapidly destroyed
by electron irradiation: Atomic resolution images of a carbon material
or molecule require a dose of several thousand e$^{-}/$\AA$^{2}$,
which exceeds the damage threshold of organic molecules by orders
of magnitude \cite{Egerton2012,Hovden2012}.

A promising approach to circumvent the dose limitation is to distribute
the required dose over many identical copies of a given object. This
has been the basis for numerous successful object reconstruction schemes
for biological electron microscopy \cite{Frank1978,Zhu1997,VanHeel2000,Scheres2005}.
If single objects can be identified and oriented in TEM micrographs,
it is possible to recover a high signal-to-noise ratio image by superimposing
all snapshots that correspond to the same orientation and conformation.
However, for lowest doses and/or smaller objects, the alignment is
problematic \cite{Sigworth1998} or fails. A similar problem exists
with x-ray diffraction data from individual molecules or nanocrystals
as recorded with the recently developed pulsed x-ray beams from free
electron lasers \cite{Neutze2000,Chapman2006,Bogan2008}, where the
orientation of the molecule in each snapshot is not known \textit{a priori}
\cite{Fung2008,Loh2010}. The tasks of retrieving objects either
from low-dose diffraction data or direct images are indeed closely
related. The unknown parameters comprise in the former rotations and translations and 
in the latter only rotations. Powerful statistical approaches
have been developed in order to recover the structure even when the
dose is not sufficient for a straightforward assignment of these parameters
\cite{Elser2008,Loh2009,Schwander2010,Saldin2010,Elser2011,Giannakis2012,Schwander2012,Philipp2012,Tegze2012,Kucukelbir2012,Walczak2014,Ayyer2014,Loh2014}.

In a recent work, we have considered the distribution of dose over
many identical objects for the case of a crystalline lattice decorated
with radiation-sensitive point defects or small molecular adsorbates
\cite{meyer2014atomic}. In this case, the periodicity of the underlying
lattice leads to a finite set of possible translations and rotations.
\del{In other words, it is assumed that there is a finite set of possible
deviations from the periodic lattice, e\@.g\@., defects in the lattice,
or small molecules adsorbed in one preferred way onto the lattice.}
While the previous paper \cite{meyer2014atomic} was a proof of concept
demonstration, the present work explores the ultimate limitations of this approach \add{and contains initial results from experimental data.}

In our simulated data, we consider the case of small molecules adsorbed onto a graphene sheet. The reconstruction algorithm now incorporates all the symmetries of the underlying graphene support (while still allowing for fully asymmetric molecules). Quite remarkably, this improvement results in a $\sim$ 100-fold decrease in the required dose for the maximum likelihood (ML) reconstruction. \add{Here we firstly establish the theoretical performance limits of the reconstruction algorithm with well defined simulated input data. In particular the required amount of raw data at a given signal to noise level as defined by the intrinsic standard deviation of the infinite dose smooth object in respect to the actual standard deviation in individual pixels. Further we demonstrate the capability to correctly identify and single out individual objects from a mixture and also confirm the area independent ultimate limit of three extraneous counts per entity.  Finally, we successfully test the reconstruction algorithm with experimental data of a di-vacancies in graphene, which contains a mixture of three different divacancy configurations and also a minority of undefined intermediates.}

\section{Overview}

\begin{figure*}
\mbox{%
\lblgraphics[white]{a}{\includegraphics[width=0.166\linewidth]{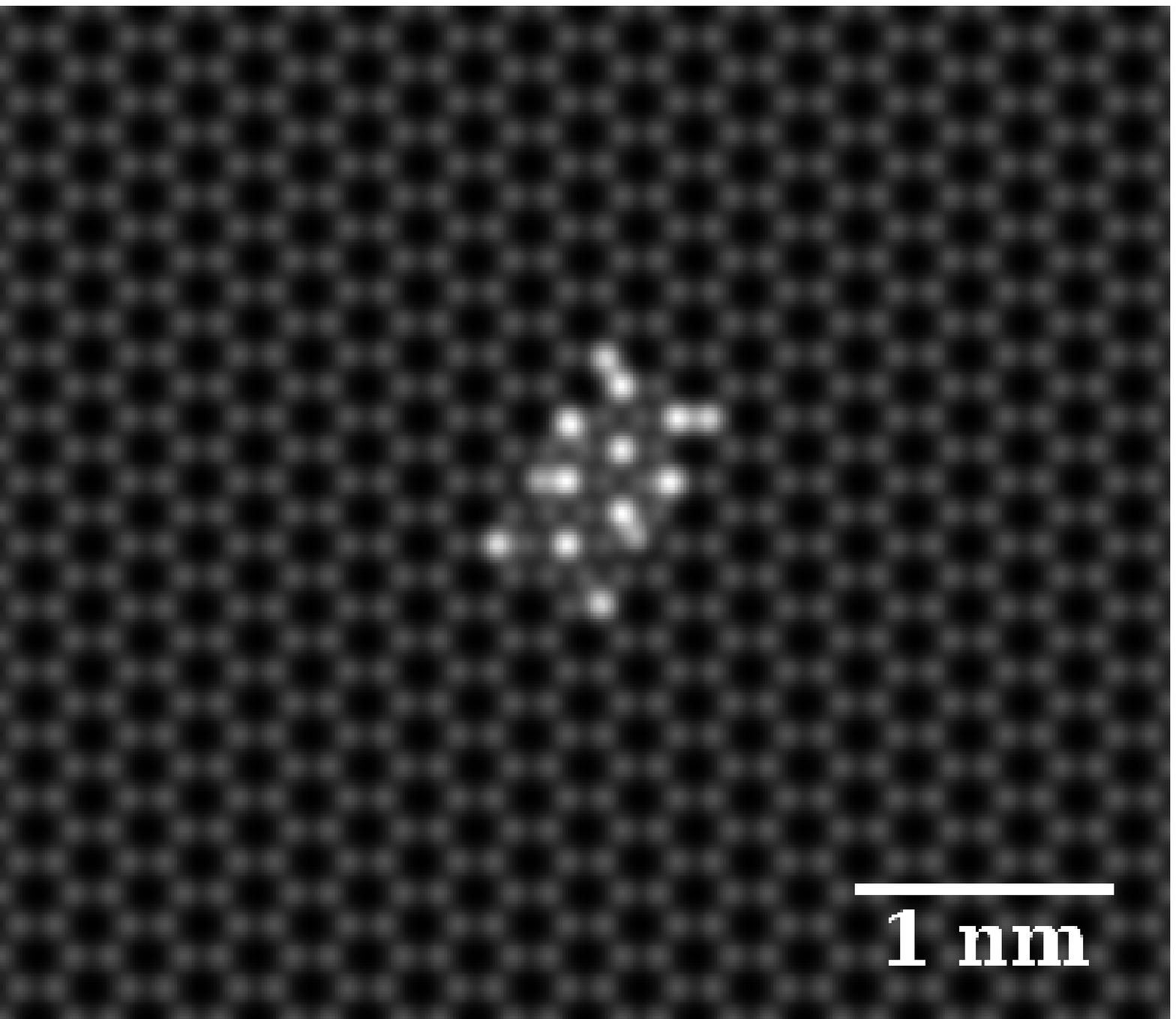}}%
\lblgraphics[white]{b}{\includegraphics[width=0.166\linewidth]{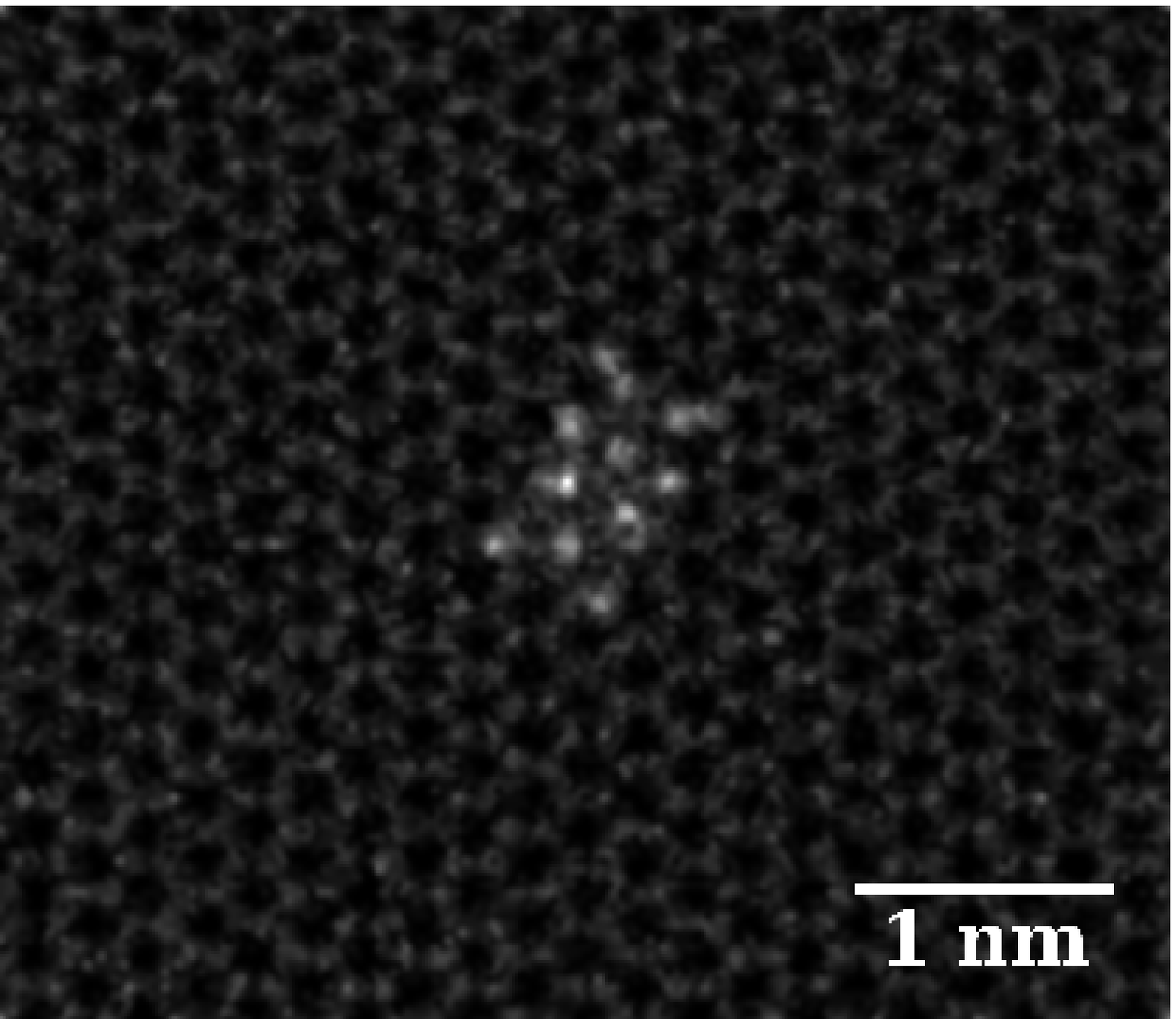}}%
\lblgraphics[white]{c}{\includegraphics[width=0.166\linewidth]{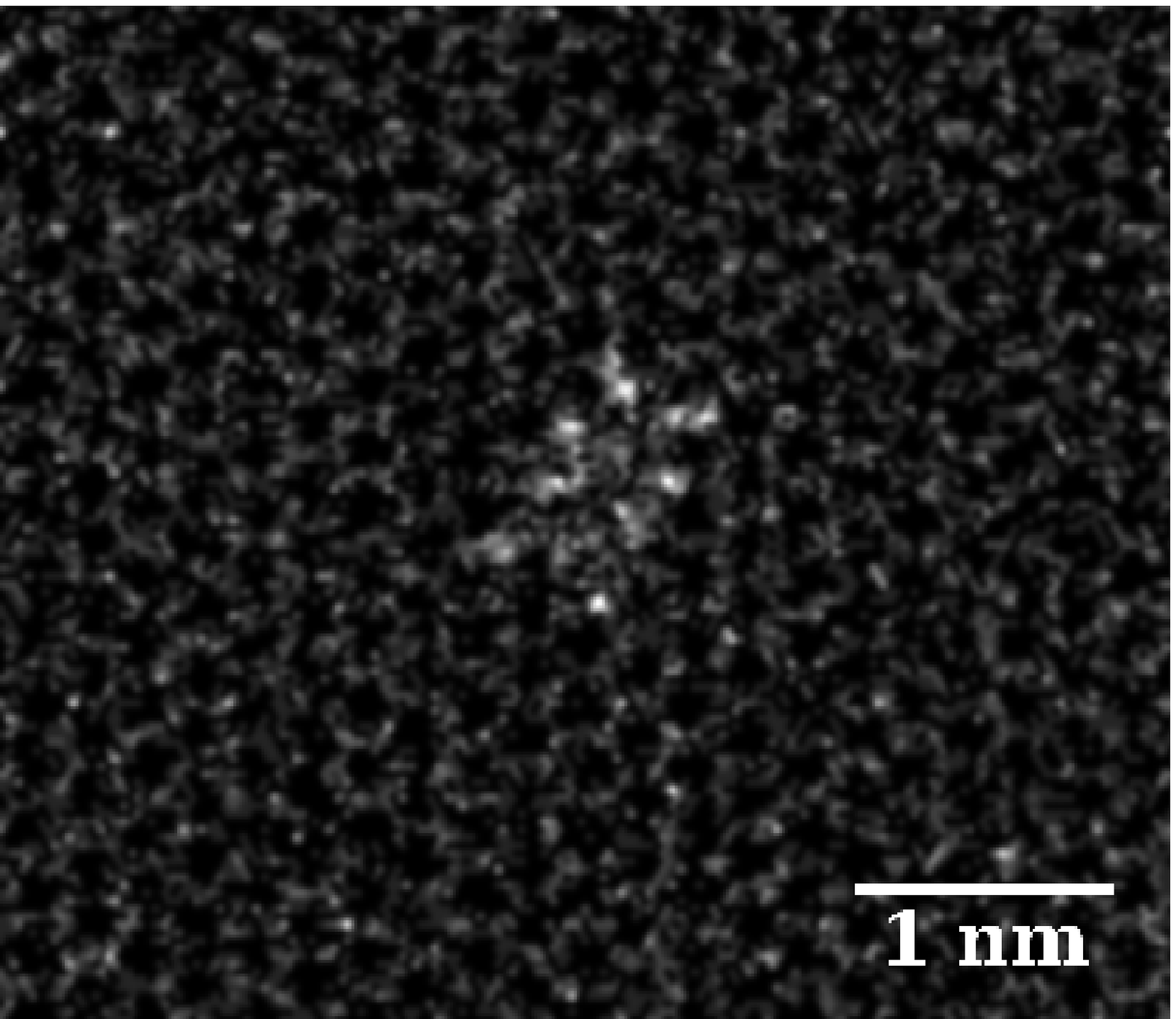}}%
\lblgraphics[white]{d}{\includegraphics[width=0.166\linewidth]{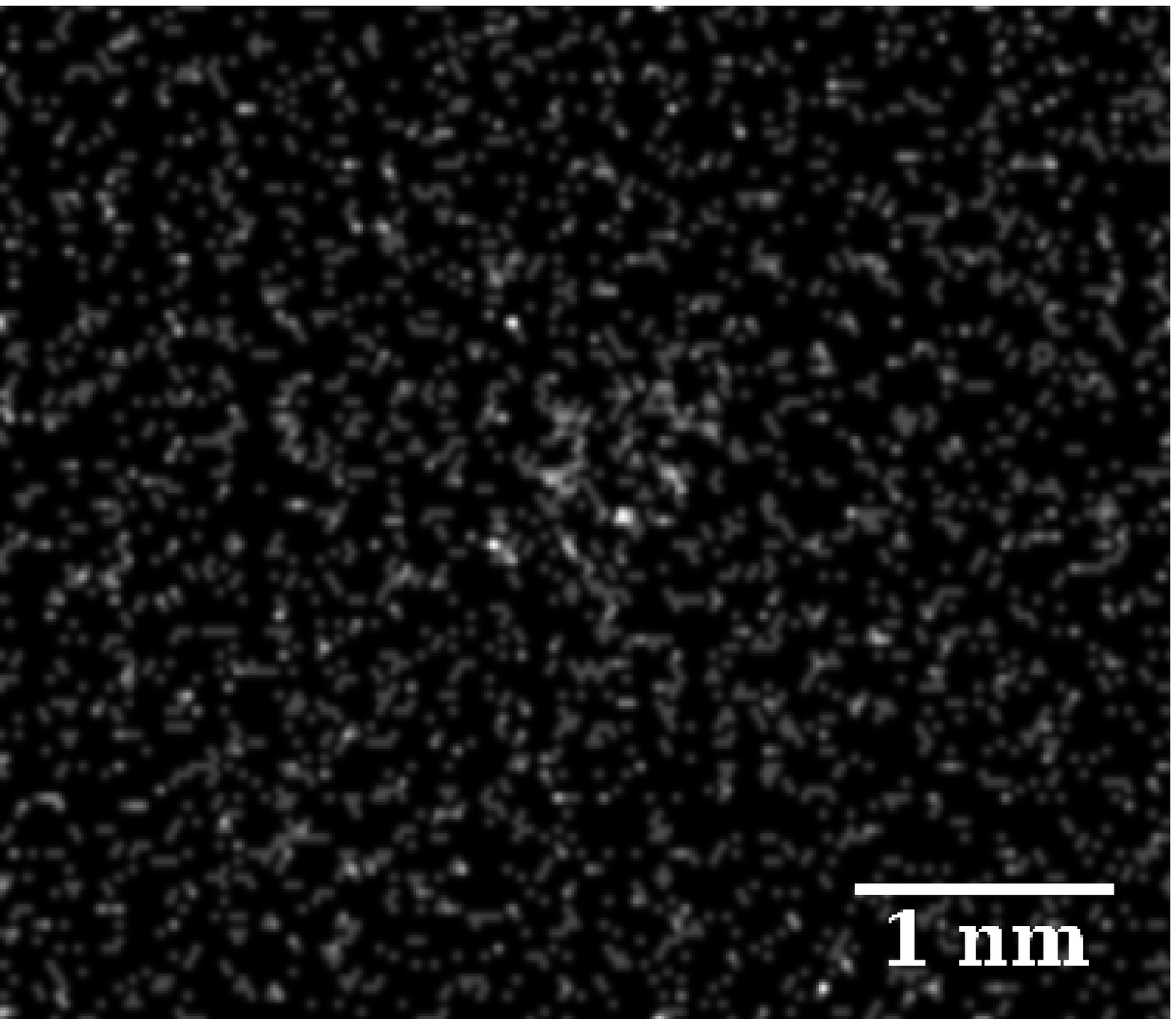}}%
\lblgraphics[white]{e}{\includegraphics[width=0.166\linewidth]{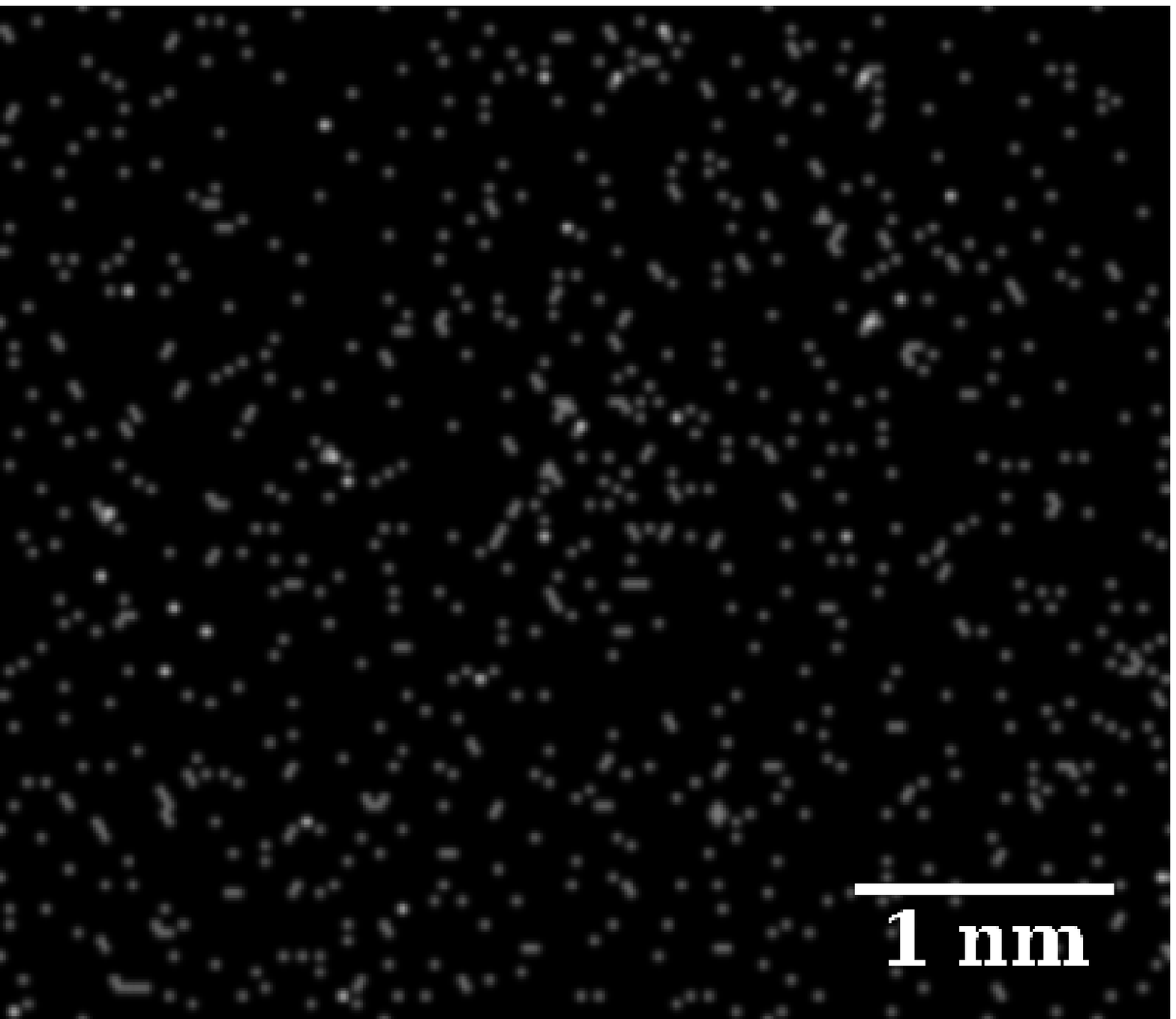}}%
\lblgraphics[white]{f}{\includegraphics[width=0.166\linewidth]{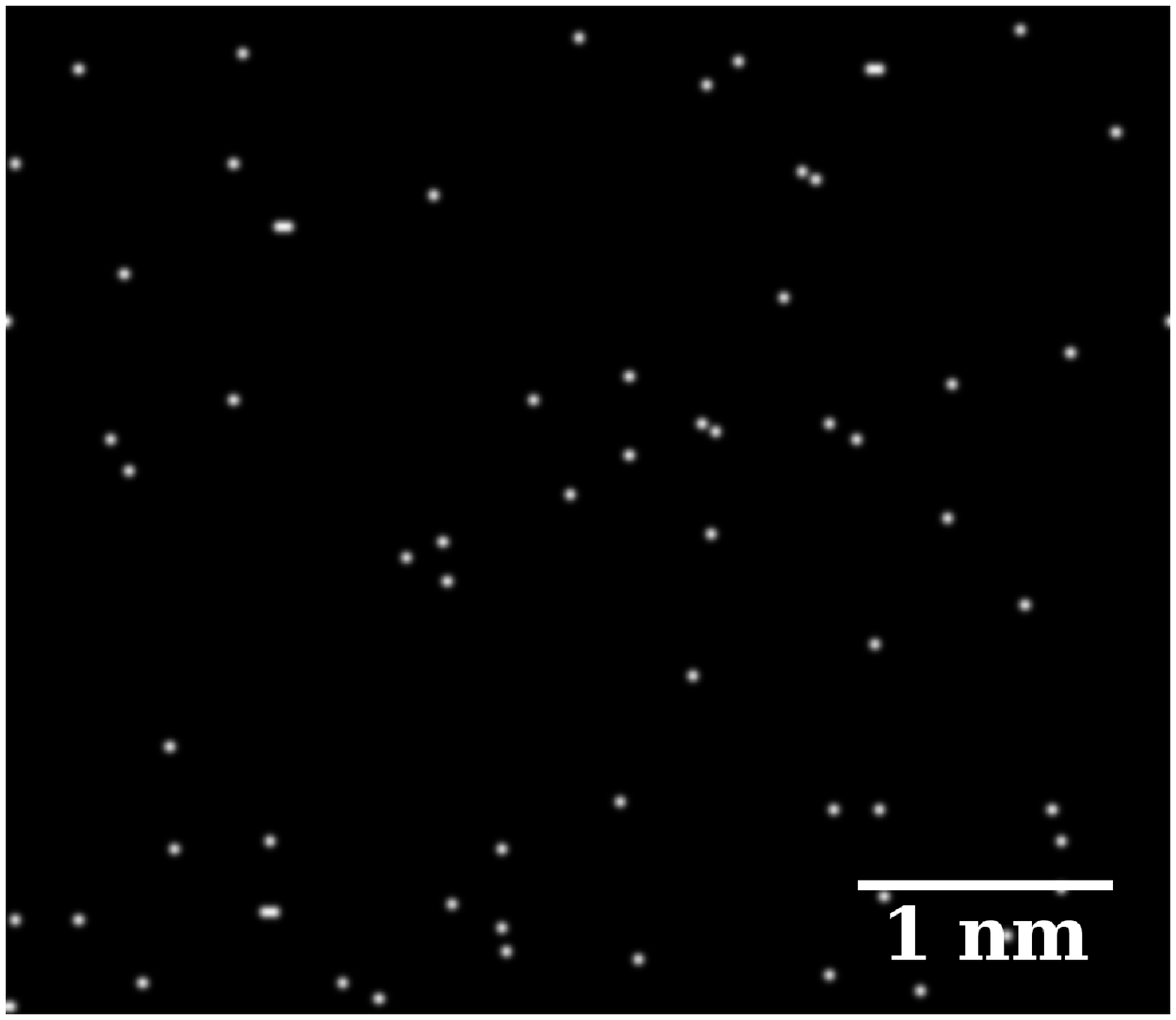}}%
}%

\begin{tabular*}{1\textwidth}{@{\extracolsep{\fill}}cccccc}%
e$^{-}/$\AA$^2$ & 2325 & 775 & 230 & 65 & 5\tabularnewline
cts\@./\AA$^2$, cts\@./Carbon & 18 , 47  & 6 , 16 & $1.8$ , 5 & $0.5$ , $1.3$ & $0.04$ , $0.1$\tabularnewline
$\sigma_{\rm gra.}/\sigma_{\rm pix.}$ dB & $-0.5$ & $-2.9$ & $-5.5$ & $-8.3$ & $-14$ \tabularnewline
\end{tabular*}%

\caption{Shot noise limited image simulations of a single F$_{4}$TCNQ on graphene
at decreasing counts or doses. Leftmost image is the noise-free simulation,
other images are calculated for the given primary dose and corresponding
counts per \AA$^2$ or per carbon atom.\del{ under medium-angle annular dark field STEM
  imaging conditions} \add{Also indicated is the signal to noise ratio, given as the contrast of graphene (standard deviation in the noise-free image) divided by the average of the standard deviations of equivalent pixels and converted to decibels.}}%
\label{fig:dose}%
\end{figure*}

For our numerical demonstration and as a suggested candidate for experimental
realization we first chose Tetrafluorotetracyanoquinodimethane (F$_{4}$TCNQ)
because it is known to exist as individual molecules or short range
ordered mono layers on graphene \cite{Gao1997}. The Flourine atoms
give additional contrast and the absence of hydrogen will eliminate
initial hydrogen stripping events from the possible degradation pathways.
Figure~\ref{fig:dose} shows simulated medium-angle annular scanning
transmission electron microscopy images of F$_{4}$TCNQ adsorbed on
a graphene membrane. The imaging conditions as chosen in the simulation
correspond to those routinely used for imaging graphene and other
2-D materials on an aberration-corrected scanning transmission electron
microscope (STEM). However, imaging the atomic structure of individual
molecules is not realistic given that the doses required for a sufficient
signal to noise ratio are orders of magnitude higher than the critical
doses of similar organic structures.

\begin{figure}
\mbox{%
\lblgraphics[white]{a}{\includegraphics[width=0.5\linewidth]{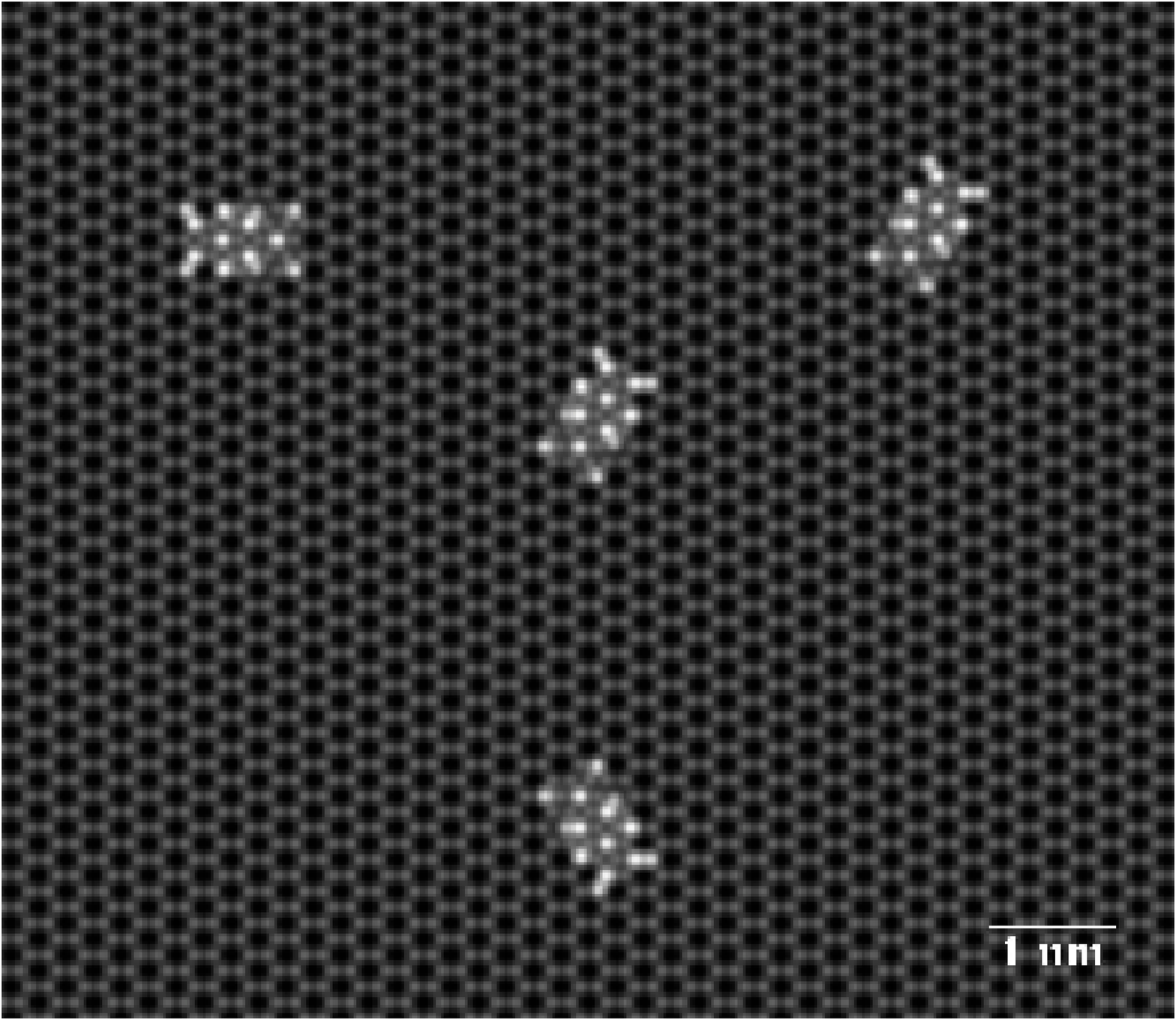}}%
\lblgraphics[white]{b}{\includegraphics[width=0.5\linewidth]{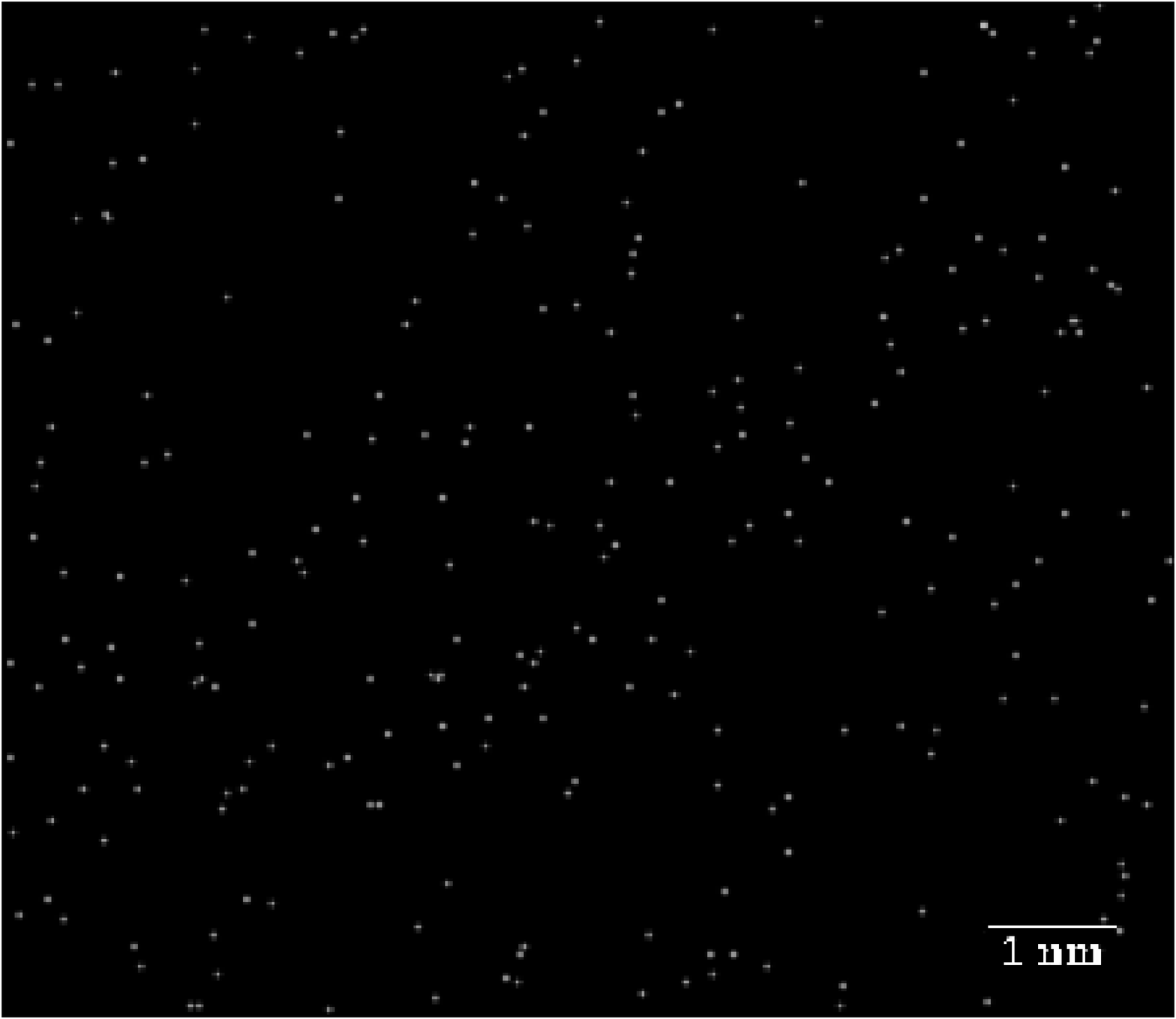}}%
}\\%
\mbox{%
\lblgraphics[white]{c}{\includegraphics[width=1\linewidth]{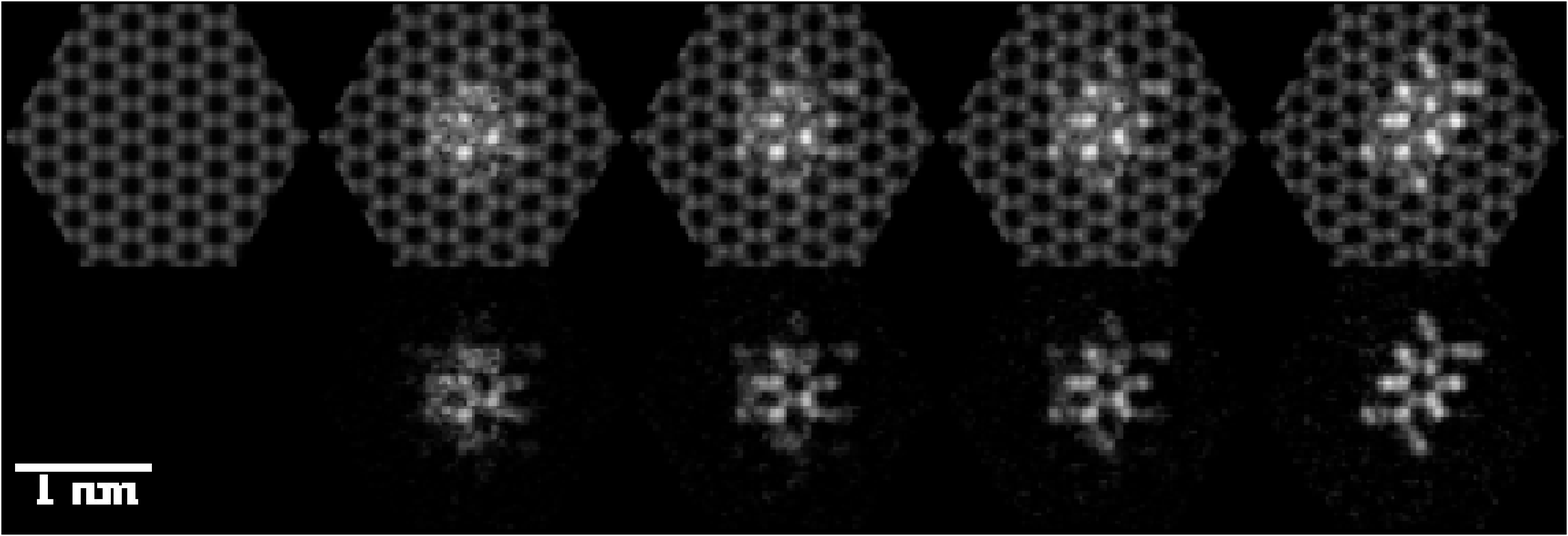}}%
}%
\caption{\textit{(a)} $71.5$~nm$^{2}$ of noise-free data with four F$_{4}$TCNQ
molecules, and (b) the same area at $0.1$ counts per Carbon (structure
not discernible). (c) reconstruction of F$_{4}$TCNQ on graphene,
most left is the input and the following images show the evolution
of the converging model image and its difference to the empty lattice.}%
\label{fig:reconstruction}%
\end{figure}

Now, we consider the case of many, randomly distributed identical
molecules on a large sample area. If we assume that the molecule adsorbs
in one specific relation to the graphene lattice, there are still 12 different
rotated and mirrored possible projections. For example, three different
projected views are contained in fig\@.~\ref{fig:reconstruction}a,
where four randomly placed molecules are visible (two of them have
an identical orientation).%

Figure~\ref{fig:reconstruction}a shows part of a large data set
with a total simulated area of 7.0~$\mu{\rm m^{2}}$ containing molecules
at a density of 0.06~nm$^{-2}$ in random, non-overlapping positions.
This data set is now processed to simulate low-dose imaging of the
entire area, with a dose of 5~e$^{-}/$\AA$^2$, resulting
in $0.04$~counts/\AA$^2$ (3.2 counts per molecule) on the
detector \add{(all simulated data assumes an ideal detector where every scattered electron generates one count, and hence the noise is the shot noise from the finite number of electrons).} Figure~\ref{fig:reconstruction}b shows the same area as fig\@.~\ref{fig:reconstruction}a
at these conditions. This is a most challenging scenario, where the
ML algorithm is put to the test of retrieving the hidden F$_{4}$TCNQ
structure

For the following analysis, the data is tiled with $F$ hexagonal
8x8 super cells with an area of $3.35$~nm$^2$. This size
is chosen to be somewhat larger than the molecule. These super cells
(or frames) will be indexed by $f=1...F$. A set of model structures
of the same size is initialized with the empty graphene lattice. We
use $M$ models, indexed by $m=1...M$. The index $s=1...S$ runs over the group of
lattice symmetry operations \add{(translation, rotation, mirroring),} and
$i=1...I$ runs through all pixels of a frame or model. Now, we calculate
(and later maximize) a likelihood value, 

\begin{eqnarray}
L & = & \prod_{f=1}^{F}\sum_{m=1}^{M}\frac{w_{m}}{S}\sum_{s=1}^{S}P_{m,f,s}\label{eqn:likelihood}\\
P_{m,f,s} & = & \prod_{i=1}^{I}P(k_{f,i},\lambda_{m,s(i)})
\end{eqnarray}

which expresses the likelihood of obtaining our snapshots under the
assumption of the model images. In the above equation, $k_{f,i}$
denote the raw data values of every frame $f$ in each pixel $i$
and $\lambda_{m,s(i)}$ are the expectation values of every model
$m$ in each pixel $s(i)$ under a lattice symmetry operation $s$.
The models are weighted with $w_{m}$. Every $P_{m,f,s}$ is the probability
to observe a frame $f$ for a given model $m$ under a symmetry operation
$s$. $P(k,\lambda)$ is the probability to observe $k$ counts for
a corresponding expectation value $\lambda$. Here we choose Poison
statistics, 
\begin{equation}
P(k,\lambda)=\frac{\lambda^{k}e^{-\lambda}}{k!}\label{eqn:poisson}
\end{equation}

but we point out that the entire reconstruction algorithm can be run
with any other probability distribution function (PDF), in particular
also the empirical histograms that can be collected from equivalent
pixels.\del{in an experimental data set} The symmetric likelihood is invariant
under permutations of the frames and the models as well as lattice
symmetry operations on any of the models or frames.
The model images and their weights are now adjusted so as to maximize
$L$. As a result, we obtain what corresponds to a single high dose
image of the molecule. A few selected steps of this optimization are
shown in fig\@.~\ref{fig:reconstruction}c. For this specific example,
the reconstruction was launched with 4 model images and the displayed
model converged towards F$_{4}$TCNQ on graphene. The other models
only show empty graphene. Evidently, the reconstruction can be classified
as successful, as the structure of F$_{4}$TCNQ is clearly visible
in the final step.

One major improvement over our earlier work \cite{meyer2014atomic}
is that rotation and inversion symmetries are now incorporated into
the likelihood function, and treated in an equal way with lattice
translations. This is realized by hexagonal pixels and super cells,
as described below. Moreover, we have explored different optimization
methods in order to find the maximum in $L$. For the example in fig\@.~
\ref{fig:reconstruction}, we find that we can reconstruct the structure
of the molecule from MAADF-STEM data with a dose of only 5~e$^{-}/$\AA$^2$,
from a total area of several $\mu$m$^2$. 

\del{%
The results
presented here are obtained with simulated data. Finite aberrations
in experimental micrographs will effectively reduce the number of
useable rotations and mirrors, but will always preserve the translational
symmteries. The experimental realization will involve technical challenges
in preparing suited, clean samples with carefully deposited molecules
as well as the automated acquisition of data from large areas.
}%

\section{Implementation}

\begin{figure}
\mbox{%
\lblgraphics[white]{a}{\includegraphics[width=0.5\columnwidth]{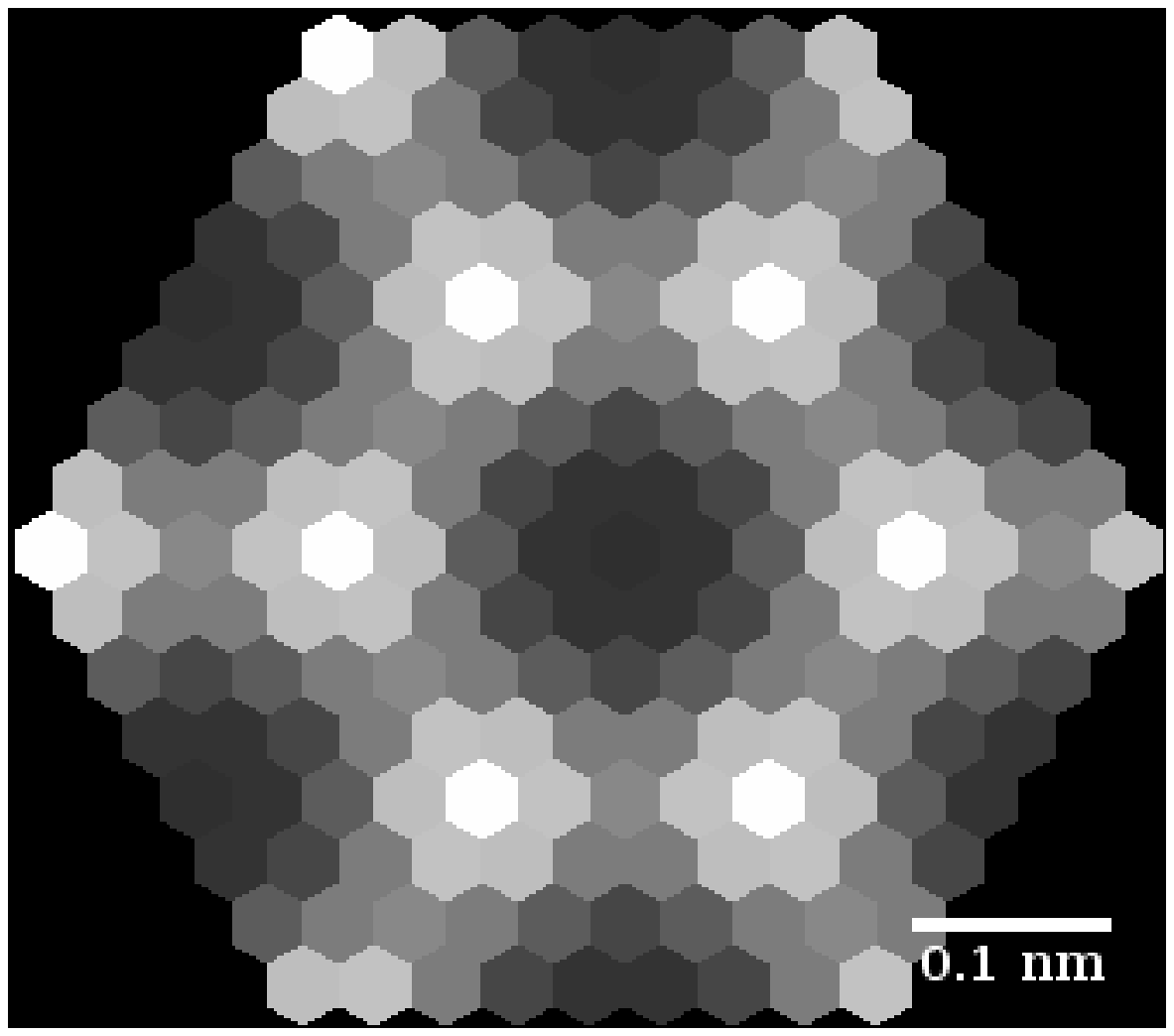}}%
\lblgraphics[white]{b}{\includegraphics[width=0.5\columnwidth]{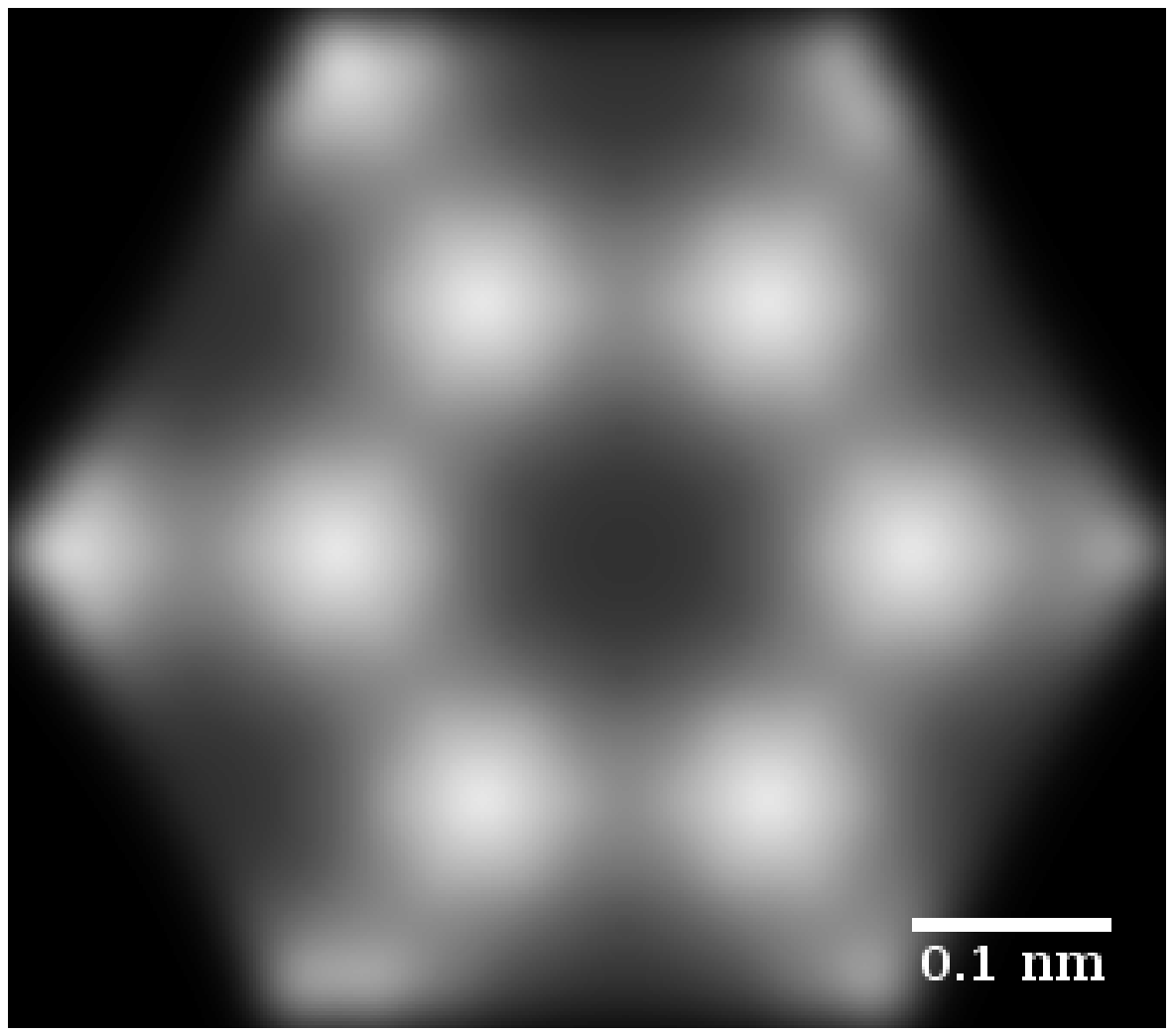}}%
}

\caption{(a) A hexagonal 2x2 supercell showing graphene with a sampling
density of 4 hexagonal pixels per bondlength ($1.42$~\AA).
In the present work we use 8x8 hexagonal super cells with an total
area of $3.35$~nm$^2$. (b) The same image
smoothed with a Gaussian FWHM of $0.32$~\AA.}%
\label{fig:frame}
\end{figure}

In order to make use of the full symmetry of the graphene support,
the image data has to be re-sampled into hexagonal pixels and
cut into hexagonal super cells or frames. This requires discernible
Bragg spots in the Fourier transformed micrographs. For a hexagonal
super cell the combinations of $60^{\circ}$ rotations, translations
and mirrors do not scramble the data when applying periodic boundary
conditions, as they inevitably would in the case of rectangular super cells.
In order to represent atomic resolution of $\sim1.0$~\AA\, at
full width half maximum (FWHM) 4 hexagonal pixels are sampled per
C-C bondlength. \add{Their edge-to-edge diameter and area are hence $0.355$~\AA\, and $0.11$~\AA$^2$, respectively.}
A close up of a 2x2 super cell is presented in fig\@.~\ref{fig:frame}.
The supercell has three longer and three shorter edges to match the
threefold periodic boundary conditions with complete pixels.

For generating image data we first simulate STEM images of graphene
with randomly seeded molecules. The molecules are assumed to adsorb
in registry with the lattice in random symmetry-equivalent configurations.
For each simulated snapshot, we render a hexagonal super cell with double
the intended diameter. Then the central region with the desired size
is cropped before the frames are noisified to the specified count
rate using Poisson statistics. The periodic boundary conditions necessitate
that the frames and models have at least double the \del{expected} size
of the molecule.

The reconstruction is initialized with a set of identical models,
that are representing the expectation values of the graphene lattice.
We use 64 gray levels to quantize the expectation values in the models.
A gray value of 0 corresponds to the mean of the hollow sites in the
graphene lattice and a gray value of 8 corresponds to the global mean
of the raw data. The initial weights are in principle irrelevant yet
practically decisive for the speed of convergence. We find that $\sim$10-
to $\sim$100-fold variations in the weights work best as models with
smaller weights adapt to accommodate rare cases more quickly. In order
to compute the vastly differing probabilities in eq\@.~\ref{eqn:likelihood}
numerically we use an offset logarithmic likelihood and treat the
mantissa and exponent separately.

For the optimization, i\@.e\@., for finding the maximum of the likelihood
function, different methods were tested. Specifically, we have implemented
a point test (coordinate descent) approach for single pixels and
small clusters of pixels adjusted simultaneously, as well as an adapted
expectation-maximization (EM) variant \cite{DEMPSTER1977,Loh2009}.
\add{ In a clustered 7 pixel test there is the full change in the center and half the change at its 6 direct neighbors.} If a change in the models
or their weights is found to increases the total $L,$ it is accepted,
otherwise rejected. Tests in $\lambda_{m,i}$ are repeated until
the current best gray value is determined.

For updating the models, all their pixels are picked in shuffled order
for optimization. After all models have been updated once, their weights
are adjusted. To speed up the convergence of the reconstruction, we
keep track of recent successful changes in the models, and if the vicinity
of a pixel has been stable lately, the optimization is only attempted
in 15\% of the cases. This scheme dynamically concentrates the \add{computational effort}
on interesting areas where the models are evolving. 

Calculating changes in $L$ by variations in the $w_{m}$ is straight
forward, if the results of the summation over $s$ in eq\@.~\ref{eqn:likelihood}
are stored, while variations in the $\lambda_{m,i}$ are calculated
by updating only the affected $P_{m,f,s}$ and their sums in eq\@.~\ref{eqn:likelihood}. This
computationally efficient way requires that all distinct $P_{m,f,s}$
and their sums over $s$ are kept in memory.
The required memory is proportional to the total area of the raw data and the number of models,
and it becomes the limiting factor on our cluster (all calculations
were done on a cluster made of 20 standard PCs with 16GB of RAM each).

The crucial difference between direct \add{pixel tests} and expectation
maximization (EM) is that the summation over the possible configurations
$s$ is not normalized, we therefore do not assume that all individual
frames $f$ contribute equally. Obviously, eq\@.~\ref{eqn:likelihood}
would simply become a constant expression if such an normalization
was applied.

In the standard expectation maximization approach, the updated expectation
values (i\@.e\@., the pixels of the model images) would be given as a weighted
sum 

\begin{equation}
\lambda_{m,i}'=\frac{1}{F}\sum_{f=1}^{F}C_{m,f}\sum_{s=1}^{S}P_{m,f,s}\, k_{f,s^{-1}(i)}\label{eqn:em_update_std}
\end{equation}
so that, in essence, the frames are summed with relative weights for
symmetry operations that correspond to their match to the prior model.
In our case of sparse adsorbates, this results in a situation where
only the (very dominant) underlying lattice is recovered, and the
(rare) deviations from the lattice are lost. We found that we could
overcome this problem by introducing an empirical scaling function
when adding a frame $f$ to the expectation values of the different
models $m$, incorporating a scaling factor in the form of a Lorentzian

\begin{eqnarray}
C_{m,f}' & = & \frac{2}{1+\Gamma^{2}\left(1+1/\Gamma-C_{m,f}\right)^{2}},\quad\Gamma=10\\
C_{m,f} & = & \frac{1}{\sum_{m'=1}^{M}C_{m',f}}\sum_{s=1}^{S}P_{m,f,s}\label{eqn:em}
\end{eqnarray}

With these rescaled coefficients the new expectation value $\lambda'$
for a pixel $i$ in a model $m$ is obtained according to eq\@.~\ref{eqn:em_update}

\begin{equation}
\lambda_{m,i}'=\frac{\sum_{f=1}^{F}C_{f',m}'\sum_{s=1}^{S}P_{m,f,s}\, k_{f,s^{-1}(i)}}{\sum_{f'=1}^{F}C_{f',m}'}\label{eqn:em_update}
\end{equation}

The choice for using the raising part below the first turning point
of a Lorentzian is entirely empirical, but well motivated considering
that frames with a significant inequality in the $C_{m,f}$ should
give a much higher relative contribution than the ones with rather
balanced $C_{m,f}$. For the EM method, we also have to use starting
models that are slightly different from each other, which is implemented
by adding a small amount of noise. The rescaled EM approach is computationally
more efficient than the ML algorithm, and produces slightly smoother
results (discussed further below). However, it is not as robust as
the ML algorithm, since its performance depends on the above described
rescaling and possibly on the way the models are initialized (we did
not explore this last point in detail).

\section{Results}

\subsection{Low dose imaging and figures of merit}

Figure~\ref{fig:dose} shows shot noise limited STEM image simulations
of F$_{4}$TCNQ on a graphene support at ever lower doses. Low doses
are required to minimize beam damage but also increase the noise in
the images. For STEM imaging, the optimum imaging conditions in terms
of the signal to noise ratio vs\@. dose are medium angle annular dark
field (MAADF) imaging conditions (sometimes also called low-angle
annular dark field), where the detector begins just outside of the
bright-field disc \cite{Hovden2012}. Here, we use the same MAADF
image simulation conditions as in our previous work \cite{meyer2014atomic}.
For a graphene structure, this results in ca\@. one count on the MAADF
detector per 130 primary beam electrons, which defines the shot noise dose in simulated data. \del{Using absolute counts on the graphene, rather than primary dose, makes our results independent of the actual
annular dark-field imaging conditions.} If we consider the
additional counts on the detector\textit{ per }molecule, we expect
that the results can be approximately transferred to other structures
of different size or mass. \add{Finally, using the resolution and sampling dependent contrast (= signal) to noise ratio ($\sigma_{\rm gra.}/\sigma_{\rm pix.}$) 
 as most general figure of merit (last line in   fig\@.~\ref{fig:dose}), the results can also be compared among different imaging methods and conditions.
 $\sigma_{\rm gra.}$ is the standard deviation due to actual image contrast and can be obtained from averaged image data, where the single pixel noise is effectively cancelled.  $\sigma_{\rm pix.}$ is the standard deviation of the intensity calculated across equivalent pixels (i.e. those with the same mean value), and then averaged across the image.
 At a noise level of $-0.5$~dB or $-2.9$~dB the molecule and lattice can be readily recognized in fig\@.~\ref{fig:dose}. At $-5.5$~dB the position of the molecule and presence of the lattice are barely discernible. And
 at $-8.3$~dB or $-14$~dB neither could be possible identified
 by direct imaging of a single occurrence.
}%
\del{%
Besides the show case example of F$_{4}$TCNQ, we have also tested the completely asymmetric guanine molecule, and a mixture of the amino acids guanine and cytosine.}%

\subsection{Required area}

\begin{figure}
\includegraphics[width=1\columnwidth]{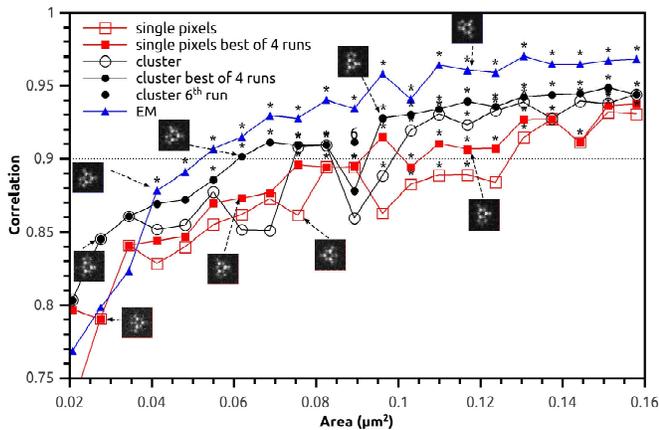}
\caption{Reconstructions of guanine from a simulated dose of 65~e$^-/$\AA$^2$
or 22 counts per molecule and a seeding density of $0.015$~nm$^{-2}$
with single pixels and clustered pixel tests for ML as well as EM. Stars
mark reconstructions that are classified as successful by visual inspection.
Insets show selected reconstructions. A correlation above $0.9$ is found
to be a close visual match.}
\label{fig:guanine_Area} 
\end{figure}

We performed systematic series of reconstructions with the aforementioned
different optimization algorithms at varying exposure area (=data
size) for the asymmetric guanine molecule on graphene. With a
dose of 65~e$^{-}/$\AA$^2$, we obtain on average 22 additional
detector counts per adsorbed guanine molecule. For each reconstruction,
the correlation between the reconstructed molecule and the noise-free
reference image is calculated. The results are presented in fig\@.~\ref{fig:guanine_Area}
where we plot the achieved correlations as a function of the used
area. We see that the rescaled EM algorithm slightly outperforms the
pixel test based ML variants. Since EM yields in essence a weighted
average of the raw-data it results in smoother reconstructions. However,
the reconstruction of sparse molecules is only achieved by a careful
choice of eq\@.~\ref{eqn:em}. Without the empirical scaling in
the $C_{f,m}'$ EM is only successful for high concentrations of the
molecule, as the contrast of the molecule does depend on its actual
concentration. We also see that for the direct ML schemes clustered
updates help to reduce the amount of required area and improve the
correlation with the smooth input molecule. Arguably this is crucial
in situations where there is only limited experimental raw data available,
yet it is computationally more efficient to generate larger sets of
raw data and test always only 1 instead of 7 pixels.

Another observation is that while success can be assured at sufficiently
large areas it becomes stochastic towards the minimally required area.
This can be overcome by repeating the reconstructions several times
with different random seeds, which is more feasible than gathering
further experimental data. The solid symbols mark the best match out
of four independent runs, while the open symbols show the results
of only the first run. As expected the maximum out of 4 runs is a
much smoother function. There is one extra data point at $0.09$~$\mu$m$^{2}$
that required 6 trials to succeed. From these results it is obvious
that the coordinate descent optimization ends up in a local maximum,
which is nevertheless a close visual match to the expected global
maximum of the likelihood function. Nonetheless, the method to find the global maximum
in $L$ deserves further consideration. With the parameters chosen
for the reconstructions in fig\@.~\ref{fig:guanine_Area}, we find
that the guanine structure could be retrieved from as little data
as $\sim$0.07~$\mu$m$^{2}$\del{under optimal conditions}.

\subsection{Mixed molecules and multiple absorption sites}

The most simple approach to resolve different absorption sites or
a mixture of different species is to launch a ML reconstruction with
multiple models, as shown for the case of vacancy-defects in graphene
in the previous paper \cite{meyer2014atomic}. Here, we use a different
approach that results in an improved performance: We begin with a
reconstruction with two model images, as if searching for a single
structure. This results in an empty lattice model and a
model that contains a mixture of the molecular species. Then, the
models are duplicated and the respective weights are split. The reconstruction
is relaunched with the doubled set of models. These steps can be repeated
until no more new structures are found.

\begin{figure}
\includegraphics[width=1\columnwidth]{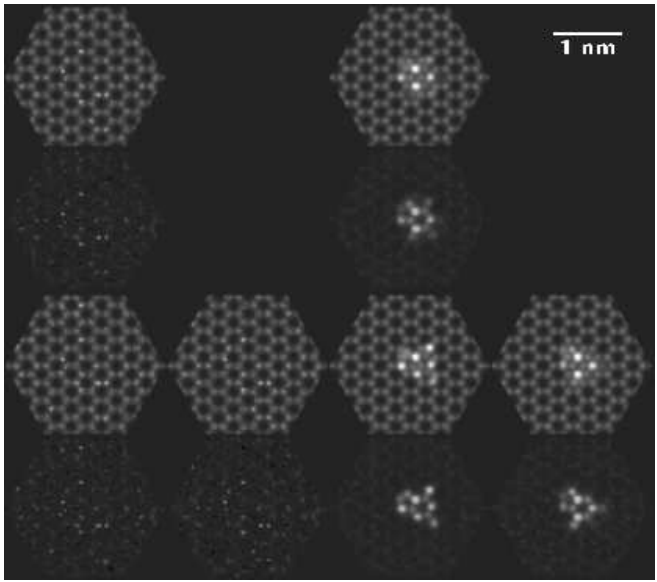}
\caption{Reconstruction of guanine on graphene in two different absorption
sites with 50\% occupancy. Rows show converged models of the first
and second run. No new structures were found in the third run (not
shown). The raw data contained $0.06$ molecules per nm$^{2}$ on a total
area of $0.22~\mu$m$^2$ at $1.2$ counts per Carbon.}
\label{fig:mixedGG} 
\end{figure}

\begin{figure}
\includegraphics[width=1\columnwidth]{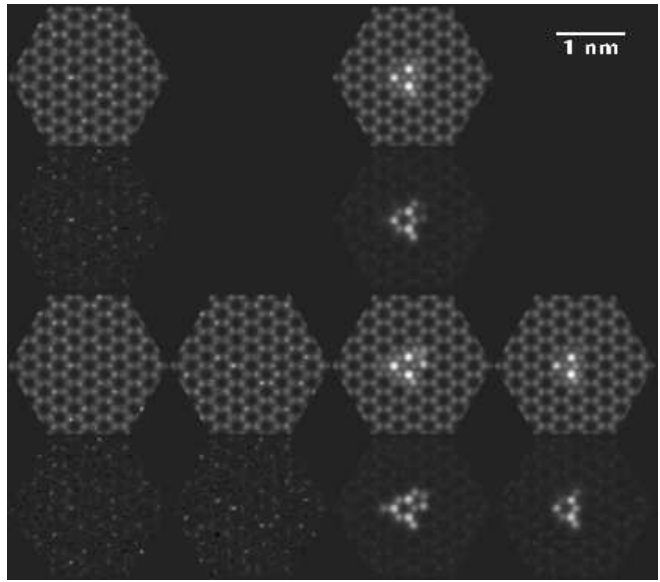}
\caption{Reconstruction of a $1:1$ mixture of guanine and cytosine. Rows show
converged models of the first and second run. No new structures were
found in the third run (not shown). The raw data contained $0.06$ molecules per
nm$^{2}$ on a total area of $0.44~\mu$m$^2$ at $1.2$ counts per
Carbon.}
\label{fig:mixedCG} 
\end{figure}

To demonstrate the capability of resolving different absorption sites
and molecules we chose two differently adsorbed guanine molecules
as well as a mixture of guanine and cytosine. Snap shots of these reconstructions
are presented in figs\@.~\ref{fig:mixedGG}\&\ref{fig:mixedCG}. All
pure cases may be readily reconstructed from an area of $0.11$~$\mu$m$^{2}$ with
a fixed molecular density of $0.06$~nm$^{-2}$ and a count rate of
$1.2$ per carbon in the lattice and 17 or 22 counts per cytosine
and guanine, respectively. For the two different absorption sites
of guanine we find that double the area ($0.22$~$\mu$m$^{2}$) with
half the concentration for each adsorption site of guanine is sufficient.
However, resolving a mixture of cytosine and guanine requires under
the same conditions an area of $0.44$~$\mu$m$^{2}$. This finding
can be understood since different absorption sites are anti correlated
under lattice symmetries, while different molecules in a similar stacking
are harder to discern (it was assumed that the ring of the guanine
and cytosine would adsorb on graphene in the same stacking, so that
the two molecules differ only in their side groups).

It is important to point out that reconstructions with mixed molecules
or mixed absorption sites were only successful with the pixel-based
(or pixel-cluster based) ML approach, while even the rescaled EM failed
to differentiate the different structures. From eq.~\ref{eqn:em_update_std}
or \ref{eqn:em_update}, it is evident that for EM there must be enough
signal in each frame so that the respective case ($m$ and $s$) can
be discriminated from the values in the $P_{m,f,s}$ (note that the
distinction between empty lattice and a single molecule was also possible
only by a rescaling of these). Hence, while EM is very powerful for
recovering a single hidden structure it might not be the best way
to find the maximum in $L$ in the case of a mixture of different
molecules or competing adsorption sites.

\subsection{Minimum required counts}

While the symmetries and relative contrast of the deposited molecules \add{as well as
the signal to noise level} will strongly affect the overall amount of required raw data, there
should be a universal limit in terms of counts per molecule: Occurrences
with 0 or 1 count do not provide any information, except for the overall
intensity. Two counts are in principle sufficient to convey the pair
correlation function. Thus the size of the molecule is defined, but
pairs of counts could for instance never break mirror symmetry. We
reason that only frames with 3 or more counts (on different pixels)
can contain structural information on the two dimensional projection
of an asymmetric molecule, like guanine. While there will statistically
speaking always be cases with three or more counts at any finite count
rate, the situation would be that most of the frames simply cannot
contain any useful information at all. In the lowest dose simulation
we show in fig\@.~\ref{fig:dose} the average counts per molecule
are $3.2$, which is at the verge of useful counts per molecule. There
is no reason to believe that ML should not be in principle feasible
at even lower count rates, but then only the fraction of the frames
that happen by chance to have at least three counts per molecule will
be very small.

We tested the three count limit under idealized conditions, namely
with the graphene support excluded from generating background counts.
In contrast to the previous results, this is not a realistic simulation
with current instrumentation,%
\footnote{In principle it is possible to imagine a dark-field TEM imaging condition
where the graphene lattice spots as well as the (000) beam are stopped
by an aperture, prior to detection of the image at the camera. %
} but serves to establish the fundamental limit. The generated frames
were sorted according to the number of counts from the guanine molecule.
Only one single model is used since we actually know that there is
only one kind of hidden structure.

\begin{figure}
\mbox{%
\lblgraphics[white]{a}{\includegraphics[width=0.25\columnwidth]{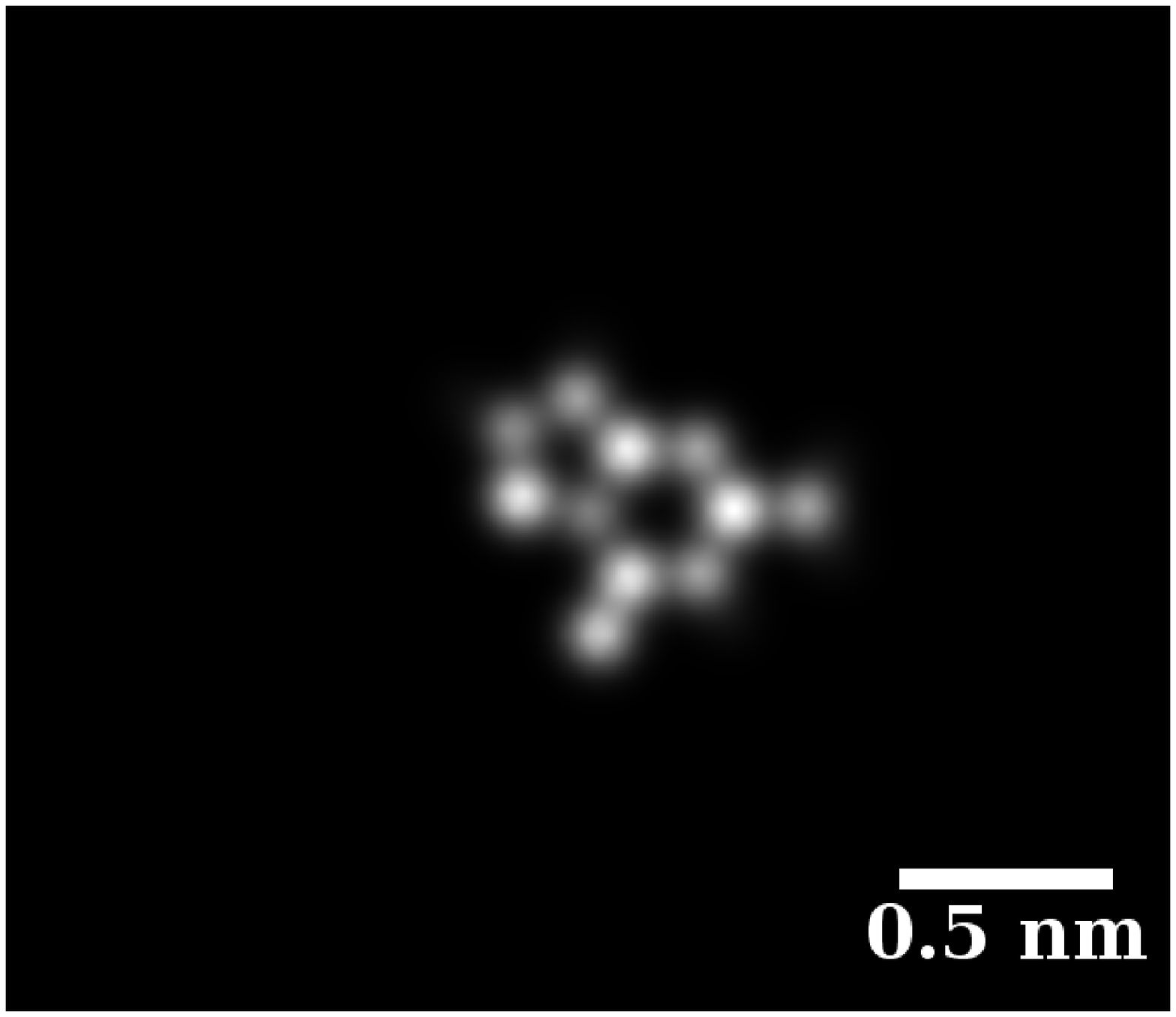}}%
\lblgraphics[white]{b}{\includegraphics[width=0.25\columnwidth]{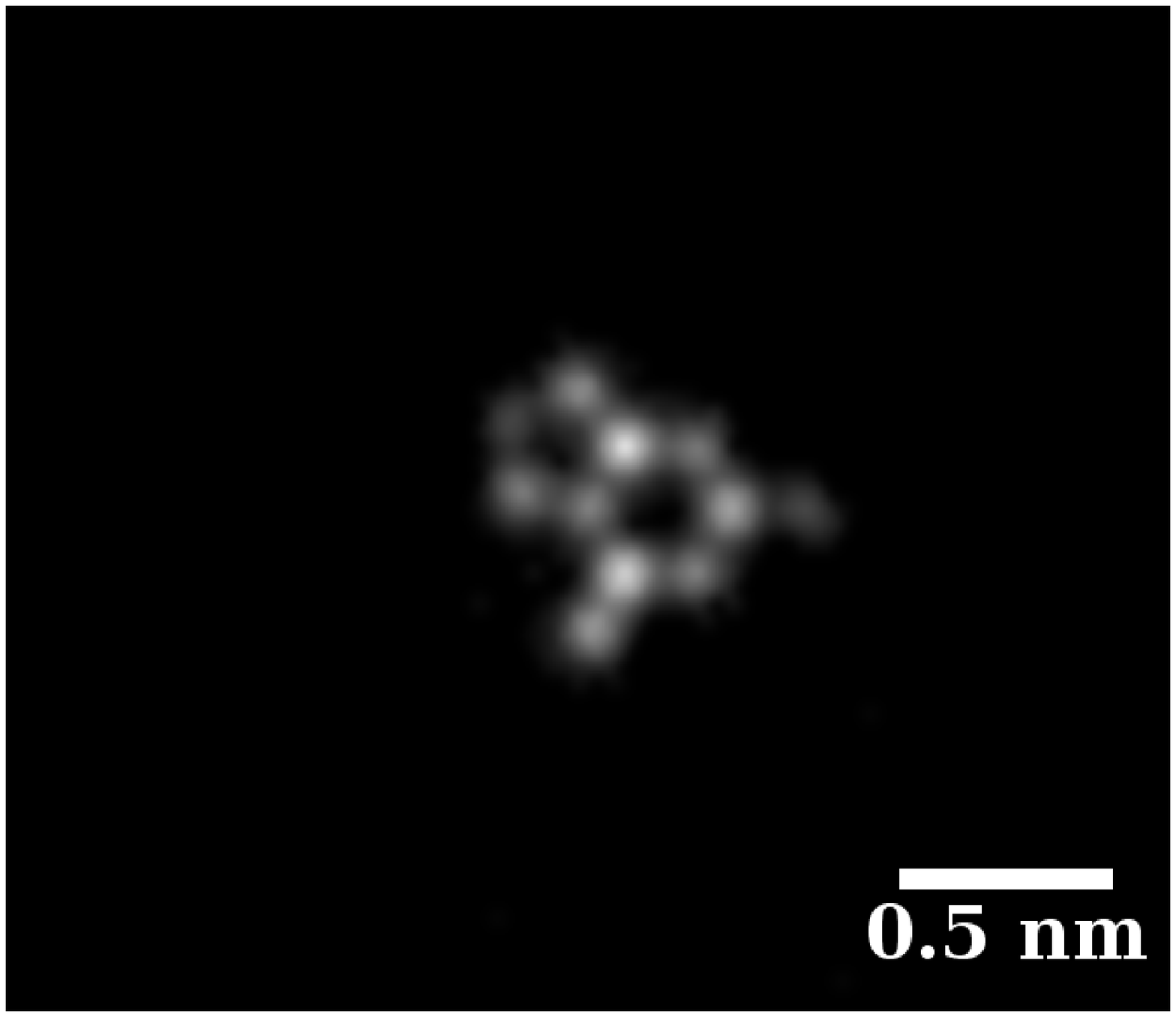}}%
\lblgraphics[white]{c}{\includegraphics[width=0.25\columnwidth]{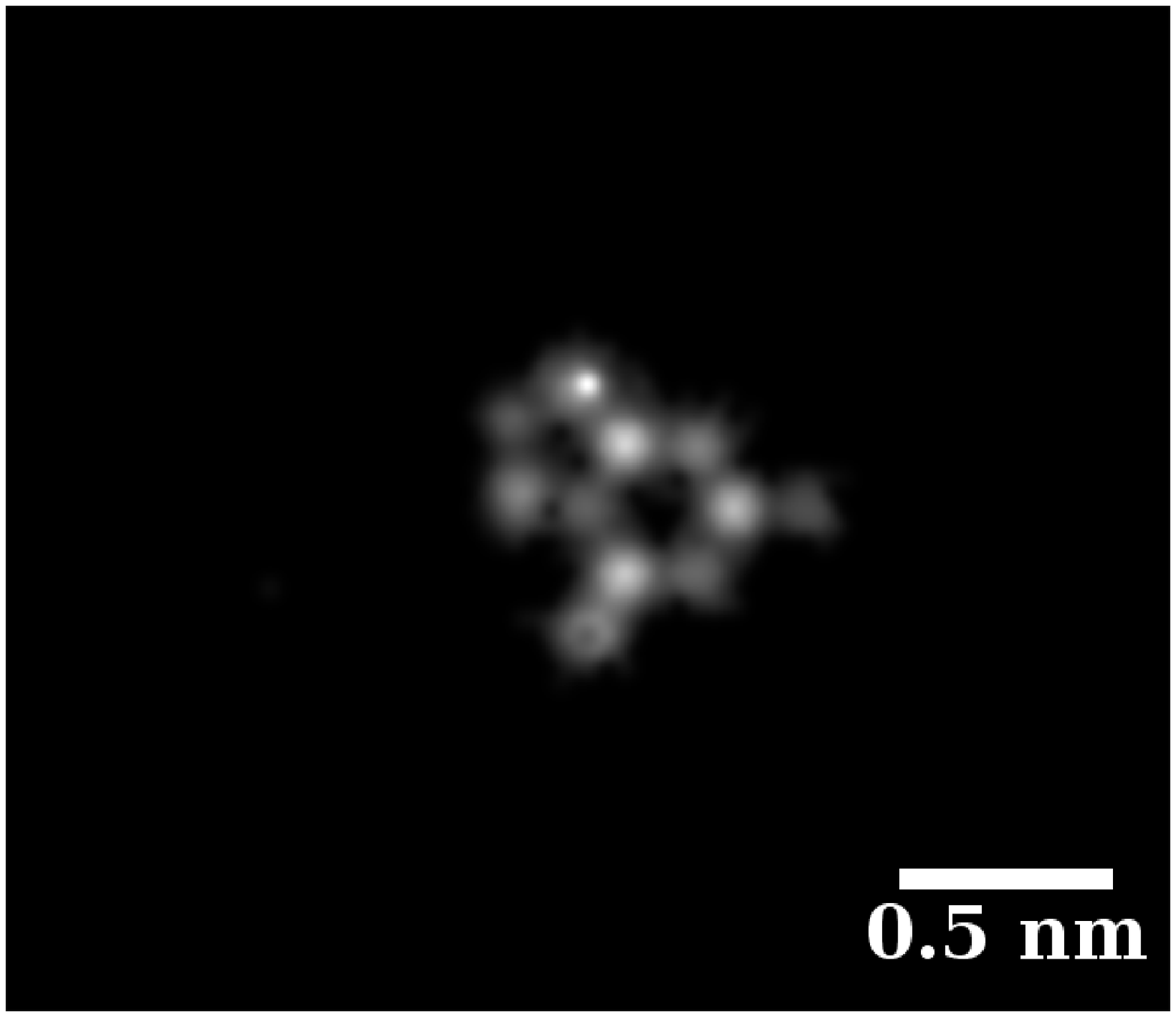}}%
\lblgraphics[white]{d}{\includegraphics[width=0.25\columnwidth]{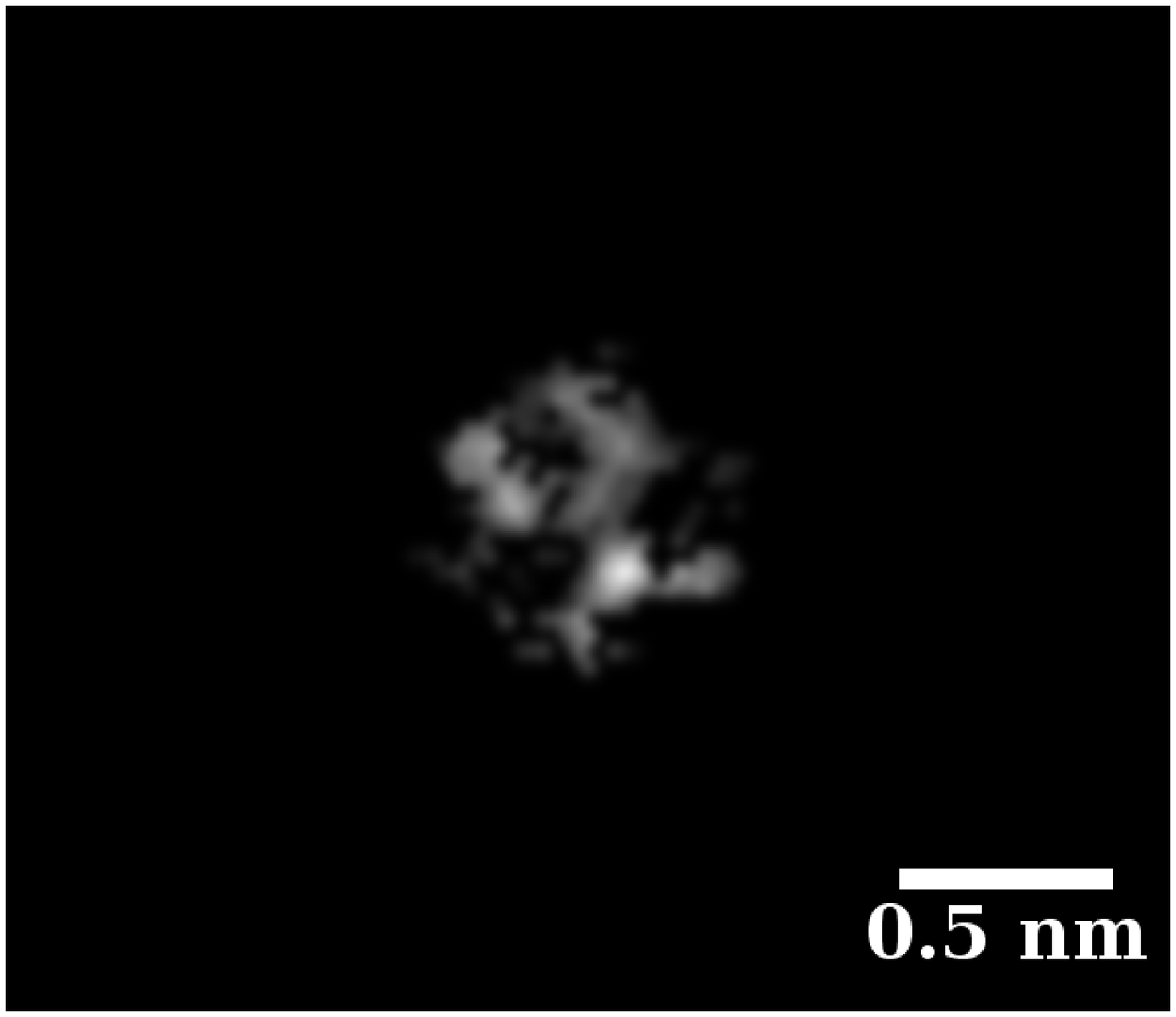}}%
}%
\caption{Reconstruction of guanine: the input (a), reconstructed
from all frames with at least one count (b), from frames with exactly
three counts (c) and from frames with exactly two counts (d).}
\label{fig:guanine} 
\end{figure}

The converged models of guanine are collated in fig\@.~\ref{fig:guanine}.
When using $4.2\times10^{6}$ non blank frames covering $14.1$~$\mu$m$^{2}$,
the guanine molecule may be readily retrieved. The reconstruction
also succeeds with the same amount of frames containing exactly three
counts. The successful reconstruction of guanine from the set of frames
with 3 counts proves that our approach still works at the theoretical
limit. It is also important to note that the reconstructions shown
in subsection B\&C (with support membrane, considered by us as realistic
simulations) also succeed at count numbers of similar magnitude. 

As expected the reconstruction of guanine cannot succeed if only frames
with exactly two counts are selected. We postulate that due to inherent
mirror symmetry in frames with exactly two counts this should be impossible
with any amount of raw data. With only 2 counts in every frame the
reconstruction does succeed for benzol or napthalene, but merely because
the \textit{a priori} built-in symmetry does suit them well. We do
not judge such special cases of high symmetry as useful for an unbiased
reconstruction of \textit{a priori} unknown and hence possibly asymmetric
molecules. 

\subsection{\add{Initial test with experimental data and treatment of aberrations}}

\begin{figure}
\makebox{\lblgraphics[white]{a}{\includegraphics[width=0.25\columnwidth]{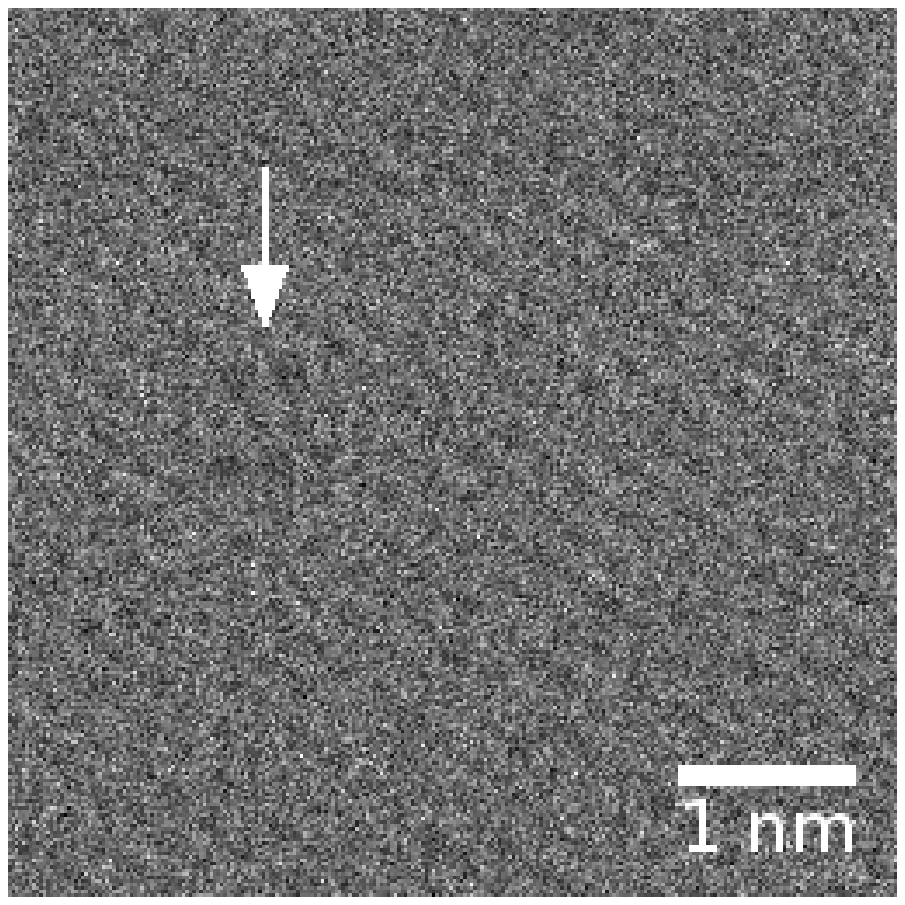}}%
\hspace{0.003\linewidth}%
\lblgraphics[black]{\hspace*{-1.5pt}b}{\includegraphics[width=0.75\columnwidth]{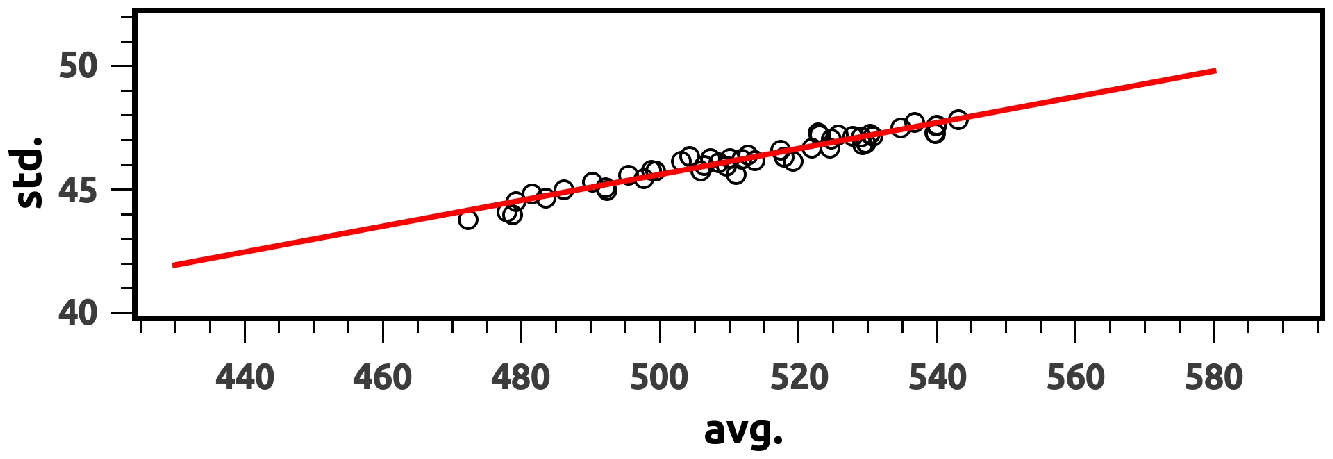}}}\\%
\makebox{\lblgraphics[white]{c}{\includegraphics[width=0.25\columnwidth]{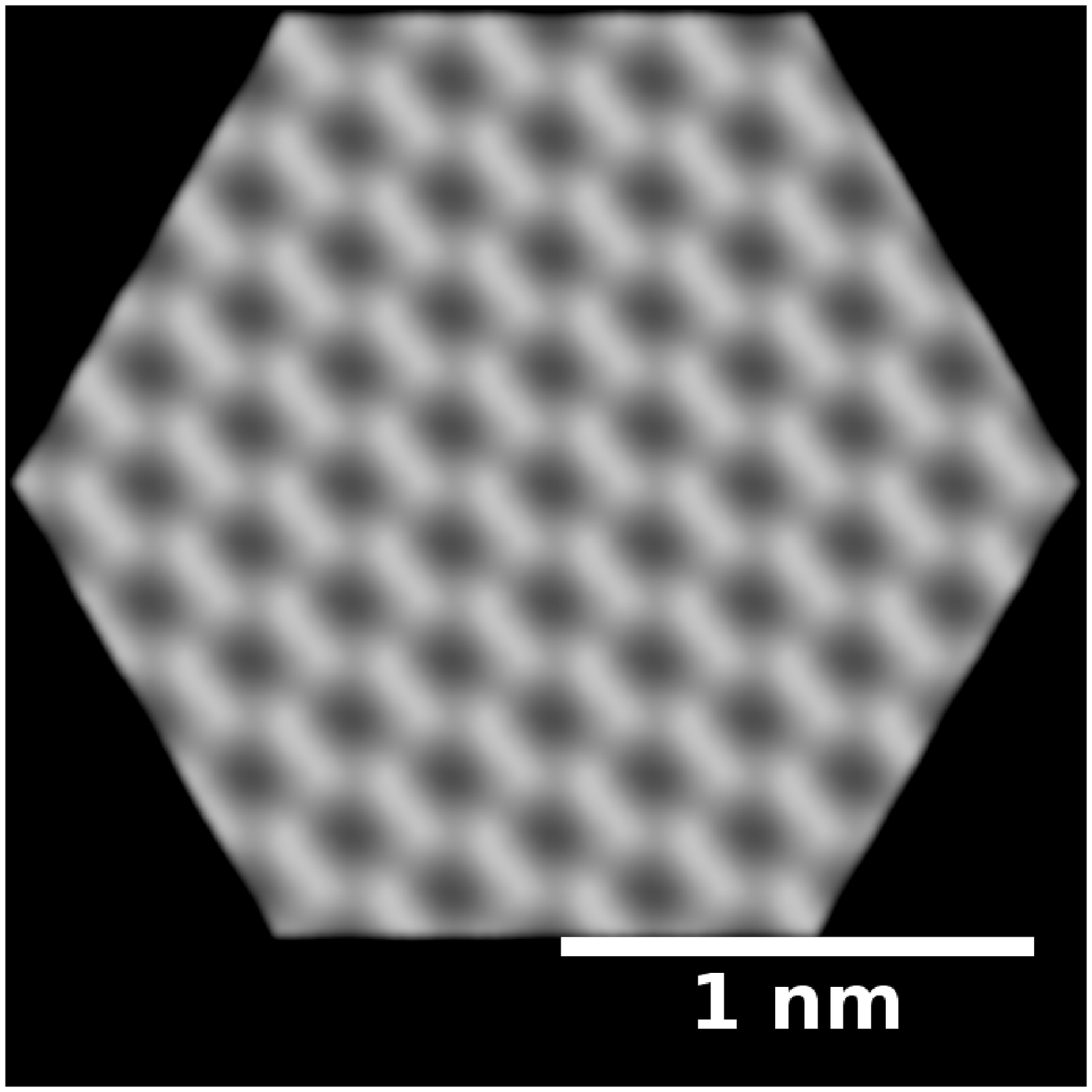}}\lblgraphics[black]{\hspace*{-1.5pt}d}{\includegraphics[width=0.75\columnwidth]{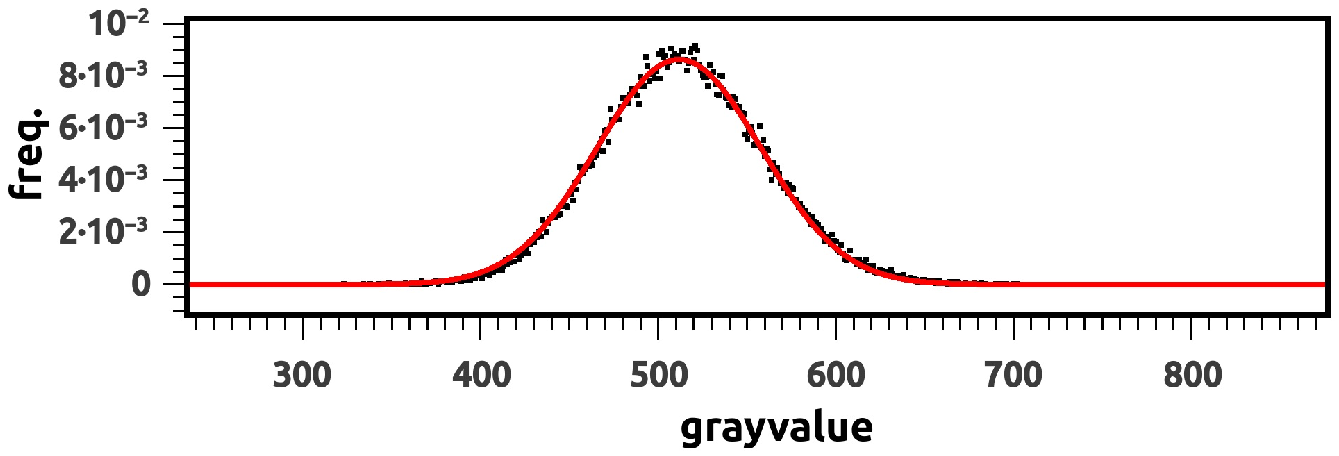}}}%

\caption{The arrow in (a) guides the eye to a 5555-6-7777 di-vacancy. The data in (b) are standard deviation vs\@. average grayvalue as obtained from the full set of 572 snapshots. The hexagonally resampled translational average (c) shows the abberated graphene lattice. The red regression line in (b) defines a Gaussian width and center that is in (d) directly compared to the distribution in each single pixel.}
\label{fig:divac2raw}
\end{figure}

For a test with experimental data we use an image sequence where a di-vacancy in graphene is rapidly switching between three configurations under the influence of the electron beam ("sequence2" from ref\@.~\cite{kotakoski2014imaging}). The original series consisted of 143 images, 512x512 pixels, with a field of view of 5~nm, and was recorded with a sufficient dose per area to visually recognize the atomic structure.  This data set was down-sampled in such a way that only 1/4th of the pixels is used for generating one new frame (by picking always one out of four pixels in each 2x2 region, we obtain a set of 572 images with 256x256 pixels).  The downsampled images corresponds to exposures with 1/4th of the original dose, and details of the structure are now at the limit of visibility by eye (fig\@.~\ref{fig:divac2raw}a).  This data set effectively corresponds to an area of $0.014$~$\mu$m$^2$. The electron dose as calculated from a beam current of $5\cdot 10^{-11}$~A and a dwell time of 16~$\mu$s is $1.4\cdot 10^5$~e$^{-}/$\AA$^{2}$. The densities of the three most common divaceny configurations are known to be $0.02$, $0.008$ and $0.006$~nm$^{-2}$\@.~\cite{kotakoski2014imaging}. There are also $0.006$~nm$^{-2}$ unclassified or corrupted occurrences. An 8x8 super cell of the translationally averaged lattice is shown in fig\@.~\ref{fig:divac2raw}c.

For our reconstruction, we have to analyze the experimental noise spectrum: Fig\@.~\ref{fig:divac2raw}b shows the correlation between standard deviation and average for all 48 individual pixels in the averaged unit cell and fig\@.~\ref{fig:divac2raw}d shows the histogram in comparison to a Gaussian according to the linear regression in fig\@.~\ref{fig:divac2raw}b. The Gaussian model and linear regression are used to extrapolate the pixel count probabilities for expectation values in the models (as replacement for eq. 3).  The signal to noise ratio, analyzed in the same way as for the simulated data in Fig. 1, is $-3.8$~dB. Hence, the experiment has a much poorer S/N ratio as expected for its dose, $+13.3$~dB would be expected under the ideal conditions of the simulations. Possible reasons are a larger inner angle of the ADF detector, a lower contrast e.g. due to residual aberrations, and additional sources of noise e.g. the background noise of the detector.

Figure~\ref{fig:divac2raw}c also reveals an astigmatism, which reduces the symmetries of the lattice. We incorporate abberations by applying a point spread function to the model images \emph{after} applying the mirror and rotation symmetry operations in eq.~\ref{eqn:likelihood}.  This point spread function (PSF) is also subject to optimization. In this way we obtain both, the defect structures (which follow the symmetry operations of the lattice) and the point spread function due to aberrations (which has a fixed orientation). The PSF is optimized by testing changes with two or three fold astigmatic or coma contributions. Radially symmetric contributions are explicitely excluded from the PSF to avoid artifical sharpening.

\begin{figure}
\mbox{\lblgraphics[white]{a}{\raisebox{25.75pt}{\includegraphics[width=0.13\columnwidth]{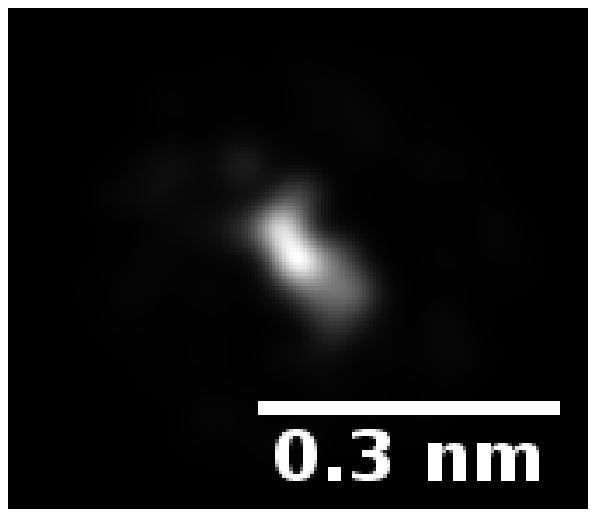}}}%
\lblgraphics[white]{b}{\includegraphics[width=0.2175\columnwidth]{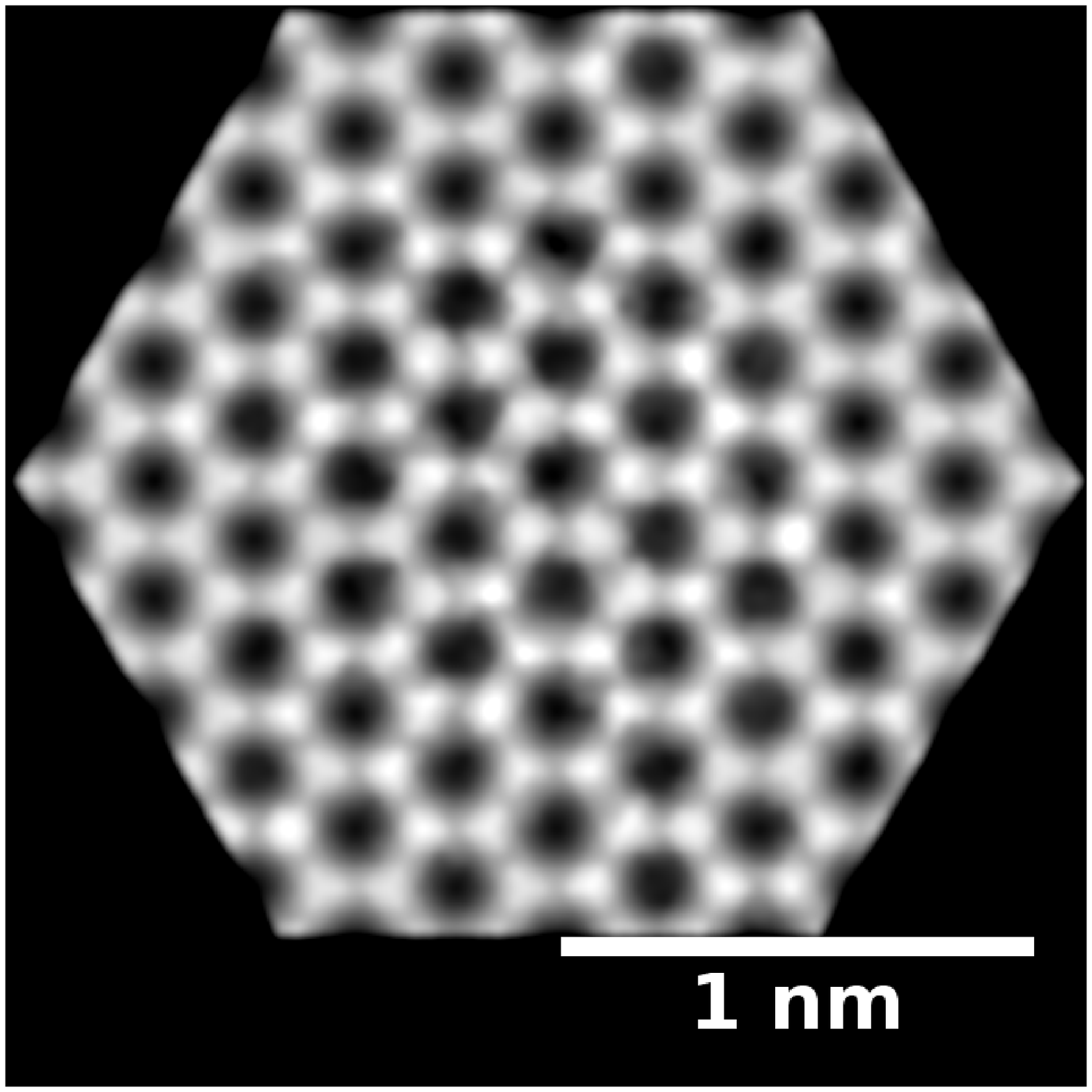}}%
\lblgraphics[white]{c}{\includegraphics[width=0.2175\columnwidth]{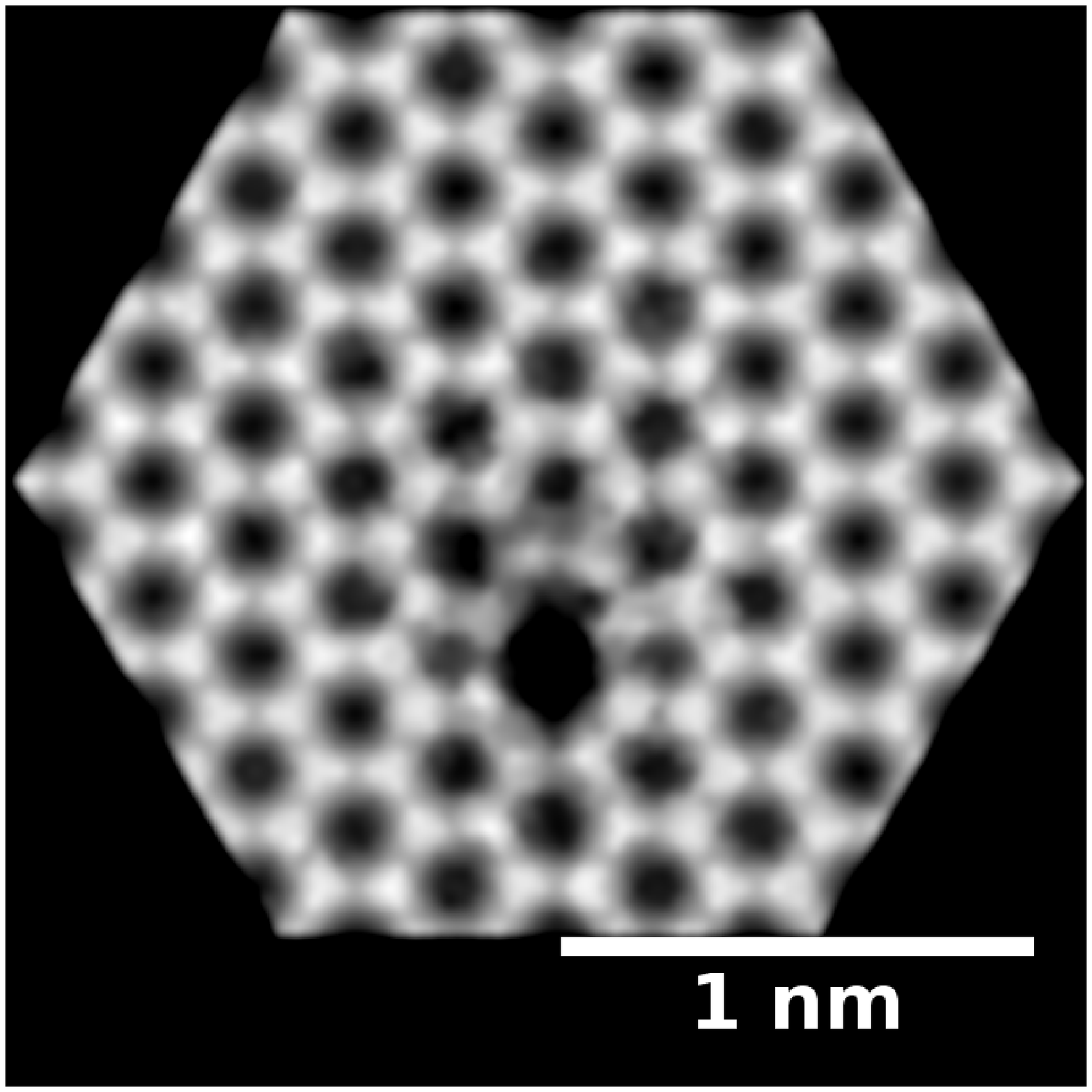}}%
\lblgraphics[white]{d}{\includegraphics[width=0.2175\columnwidth]{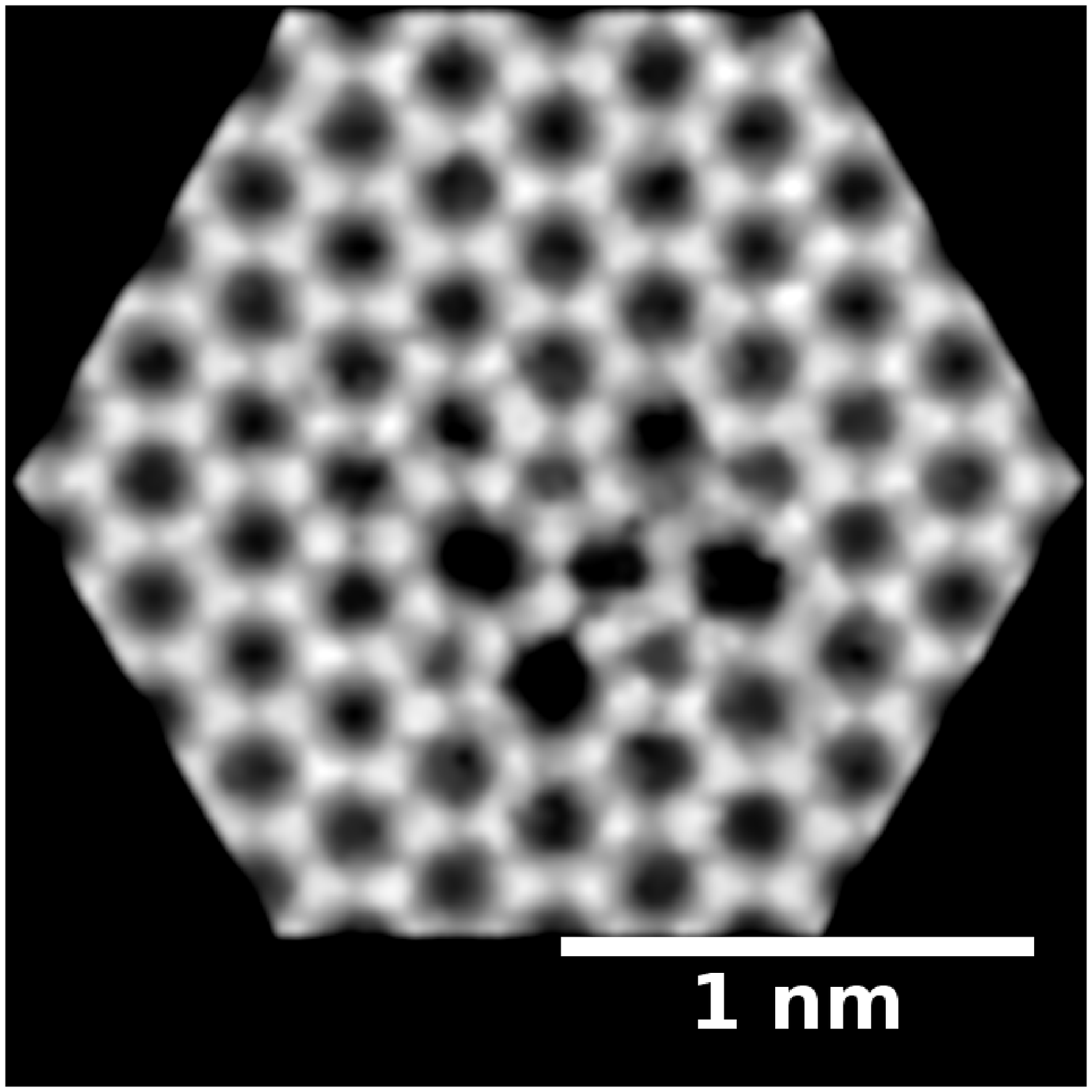}}%
\lblgraphics[white]{e}{\includegraphics[width=0.2175\columnwidth]{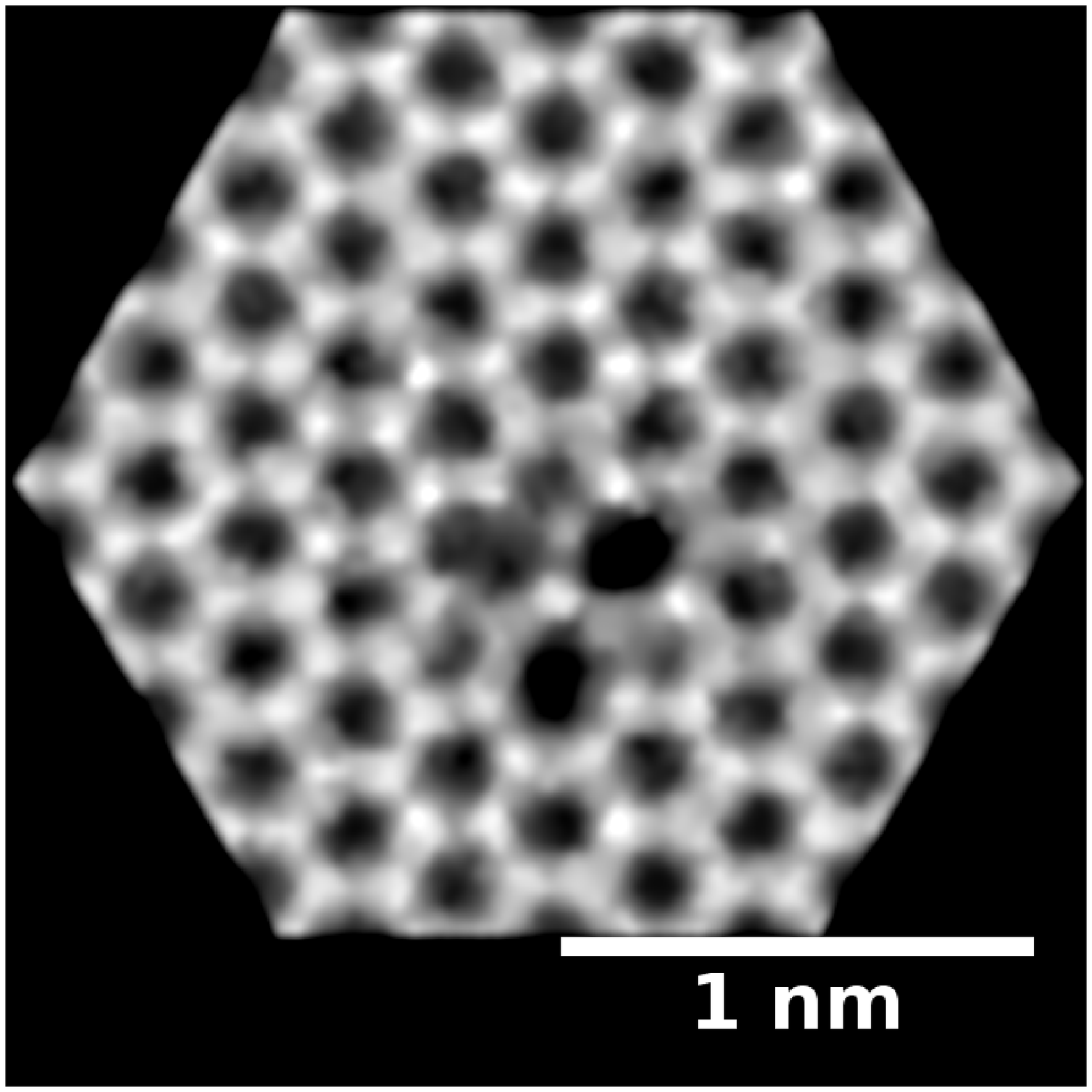}}%
}\\%

\caption{ The smaller image in (a) shows the kernel, used to represent the aberrations. The other four images (b-e) show graphene and the most prevalent divacency configurations as reconstructed from experimental data. The grayvalues in (b-e) are indentical to fig\@.~\ref{fig:divac2raw}c}
\label{fig:divac2recontstruction}
\end{figure}

 For the automated tiling by 8x8 super cells we allow an overlap of 4 lattice spacings, to more likely capture defects at least once entirely in a single super cell. In absence of boundary conditions this would ideally lead to triple over tiling, yet on a squared field of view of 5~nm this results on typically 10 partially overlapping supercells, or 5 non-overlapping super cells. 
 Figure~\ref{fig:divac2recontstruction}b shows the reconstrucion of a set of four models. The lattice as well as the three most common divacancy configurations are readily reconstructed. The inclusion of the aberration PSF does also increase the contrast in the lattice two times  as compared to fig\@.~\ref{fig:divac2raw}. The decremental effect of abberations on the image quality is therefore at least $-3~dB$. The obtained models are effectively high dose views of the most common reoccuring features in the data set, and even corrected for residual aberrations.
 As the use of the PSF can not improve the resolution, the abberation correction relies on atomic resolution in the first place, so that the PSF is defined from deviations in mirrored and rotated views.

 The optimized weights of the models correspond to defect concentrations of $0.04$, $0.015$ and $0.015$~nm$^{-2}$, which are, considering the over tiling and geometric constraints, in satisfactory agreement with the known values.
 The experimental defect densities are on the same order as the one in the simulation series in fig\@.~\ref{fig:guanine_Area}.
 
 The contrast to noise level in the experimental micrographs  is with $+4.5$~dB difference notably better than in the raw data used for the series in fig\@.~\ref{fig:guanine_Area}. In terms of data size, the recorded area is just below the range investigated in fig\@.~\ref{fig:guanine_Area}. The data in fig\@.~\ref{fig:guanine_Area} is therefore fully in line with the successful reconstruction of the most prevalent divacancy defects.  
 
\section{Discussion}

\add{We have studied the possibility to reconstruct a high-dose image from
molecular adsorbates on a crystalline support (graphene) by distributing
the dose over many identical copies of an object. This is done via
a maximum likelihood reconstruction that does not require to identify
or align the individual particles. The only fundamental prerequisite
is that there is a finite set of reoccuring deviations from the periodic lattice, in this case, molecules adsorbing in a specific way on the lattice or defect configurations within the lattice.
We have shown that the full exploit of all the symmetries of the crystalline support and assumption of a finite set of deviations are the key to retrieve the structure of adsorbed small molecules or lattice defects from low dose images.} 

There are two noticeable choices made in the definition of the likelihood
in eq\@.~\ref{eqn:likelihood}. Firstly, we sum up the individual
probabilities of all symmetry equivalent configurations. The summation
overcomes the lack of significance for a specific configuration in
a single frame, but also favors more symmetric models over less symmetric
ones. The second choice is to form a weighted sum of the models for
any given frame in eq.~\ref{eqn:likelihood}. Again, this choice
overcomes the lack of significance for even determining a best suited
model for a specific frame. 

These characteristics define the differences to other reconstruction
schemes that seek to fill-in the information gap presented by the
unknown relative configurations among the individual snapshots. Deterministic assignments as well as normalized probabilistic assessments of configurations either rely on restrictions or on sufficient statistical significance in the individual snapshots (which implies a requirement for a certain minimum dose). For example, the rescaled EM reconstruction as discussed above requires situations where the normalized $P_{m,f,s}$ \add{for a given $m$ and $f$} are significantly different for the non-matching or matching cases of $s$. In this way, an updated model is formed by matching the frames to the previous iteration. In the pixel-test based likelihood maximation, we first modify the model and then propagate the changes to all possible configurations, simultaneously maximizing the product over all frames. This reversed approach entirely circumvents the intermediate step of assigning relative orientations and translations \add{to all pairs of models and frames}. The next trial is not affected at all by a potentially noisy prior assessment of configurations. Once the
hidden structure has been retrieved, any approach should converge,
but at the low count rates investigated here our approach provides
an improvement in deriving multiple hidden structures from arbitrary
seeds in the first place.

\section{Conclusions}

\add{For the reconstructions from simulated data, we were able to reduce the required dose by a factor of $\sim100$ as compared to our earlier work \cite{meyer2014atomic} by incorporating all underlying symmetries of the graphene support membrane into the algorithm. Part of this improvement is easy to understand: In absence of the symmetry relation, the previous algorithm effectively had to
search for 12 different structures, corresponding to the same molecule
in all symmetry-equivalent positions, which are now all matched by
a single model structure. The hexagonal pixels allow to perform all symmetry operations by fast and exact integer arithmetics. Our ML algoritm is also successfully tested on mixtures of absorption sites and even mixtures of different molecules, where the required dose and/or area becomes slightly larger. Here, our new approach with subsequent optimization runs, at each stage increasing the number of models, is a powerful way to test whether one has arrived at the final result. We further
establish that the ML approach is also feasible at the fundamental limit of only three counts per asymmetric molecule. The reconstructions from the experimental rather small area divaceny micrograph series was enabled by the incorporation of full symmetries for the models and simultaneous treatment of their rotated and mirrored abberrated views - the earlier algorithm would have required an orders of magnitude larger data set.  It highlights that our ML algorithm can be successfully applied on experimental images, with real noise and in presence of aberrations.}

\begin{acknowledgments}
We acknowledge support from the European Research Council (ERC) Project
No. 336453-PICOMAT. J.~Kotakoski is acknowleged for providing the STEM image series of a divacancy in graphene \cite{kotakoski2014imaging}.
\end{acknowledgments}

\end{document}